\documentclass[12pt,preprint]{article}
\usepackage{amsmath}
\usepackage{graphics}
\usepackage{epsfig}
\renewcommand{\vec}[1]{{\bf #1}}
\usepackage{bbm}
\newcommand{\UM}{\mathbbm 1}
\newcommand{\eqb}{\begin{equation}}
\newcommand{\eqe}{\end{equation}}
\newcommand{\dmb}{\begin{displaymath}}
\newcommand{\dme}{\end{displaymath}}
\newcommand{\pd}{\partial}

\newcommand{\eab}{\begin{eqnarray}}
\newcommand{\eae}{\end{eqnarray}}
\newcommand{\ra}{\right\rangle}
\newcommand{\la}{\left\langle}

\newcommand{\be}{\begin{equation}}
\newcommand{\ee}{\end{equation}}

\begin{document}
\begin{titlepage}
\begin{flushright} 
\end{flushright}
\vspace{0.6cm}

\begin{center}
\Large{Radiative corrections to the pressure and the one-loop polarization tensor of massless modes 
in SU(2) Yang-Mills thermodynamics}

\vspace{1.5cm}

\large{Markus Schwarz$\mbox{}^\dagger$, Ralf Hofmann$\mbox{}^\dagger$, and Francesco Giacosa$\mbox{}^*$}

\end{center}
\vspace{1.5cm} 

\begin{center}
{\em $\mbox{}^\dagger$Institut f\"ur Theoretische Physik\\ 
Universit\"at Heidelberg\\ 
Philosophenweg 16\\ 
69120 Heidelberg, Germany\vspace{0.5cm}\\ 
$\mbox{}^*$Institut f\"ur Theoretische Physik\\ 
Johann Wolfgang Goethe -- Universit\"at\\ 
Max von Laue -- Str. 1\\ 
D-60438 Frankfurt am Main, Germany}
\end{center}
\vspace{1.5cm}
\begin{abstract}

We compute the one-loop polarization tensor $\Pi$ for the on-shell, massless mode in 
a thermalized SU(2) Yang-Mills 
theory being in its deconfining phase. Postulating that 
SU(2)$_{\tiny\mbox{CMB}}\stackrel{\tiny\mbox{today}}=$U(1)$_Y$, we 
discuss $\Pi$'s effect on the low-momentum part of 
the black-body spectrum at temperatures 
$\sim 2\cdots 4$\,$T_{\tiny\mbox{CMB}}$ where 
$T_{\tiny\mbox{CMB}}\sim 2.73\,$K. A table-top experiment 
is proposed to test the above postulate. As an application, 
we point out a possible connection with the stability of dilute, cold, and old 
innergalactic atomic hydrogen clouds. We also compute the two-loop correction to 
the pressure arising from the instantaneous massless mode 
in unitary-Coulomb gauge, which formerly was neglected, 
and present improved estimates for subdominant corrections.

\end{abstract} 

\end{titlepage}

\section{Introduction}

In \cite{Hofmann2005} a nonperturbative approach to SU(2) and SU(3) 
Yang-Mills thermodynamics was put forward. Here we are only concerned 
with the deconfining phase for the case SU(2). 

Briefly speaking, 
the idea on how to attack strongly interacting, thermalized SU(2) 
gauge dynamics in an analytical way is as follows: At high temperatures $T$ a ground 
state is generated out of interacting calorons and 
anticalorons. Since, on the microscopic level, there is no analytical access to 
this highly complex dynamical situation an average over 
all space involving a noninteracting 
caloron-anticaloron system of trivial holonomy \cite{HS1977} is 
performed to derive the phase $\frac{\phi}{|\phi|}$ of a macroscopic adjoint scalar 
field $\phi$. Subsequently, the existence of a Yang-Mills scale $\Lambda$ 
is assumed. By fixing the modulus $|\phi|$, $\Lambda$ and $T$ together determine how large a 
finite spatial volume needs to be to saturate 
the above infinite-volume average. Next, one observes that the field 
$\phi$ neither fluctuates quantum mechanically nor statistically. 
To complete the analysis of the ground-state dynamics the 
Yang-Mills equations\footnote{Apart 
from the term describing the interaction with $\phi$ the action for the 
topologically trivial sector after spatial coarse-graining 
looks the same as the fundamental Yang-Mills action. 
The ultimate reason for this is the perturbative renormalizability 
of Yang-Mills theory \cite{'t HooftVeltmann}.} for the (coarse-grained) gauge fields in  
the topologically trivial sector is solved subject to a 
source term provided by $\phi$: A pure-gauge solution exists which shifts 
the vanishing energy density and pressure due to the 
noninteracting, BPS saturated caloron and anticaloron to 
finite values, $P^{gs}=-\rho^{gs}=-4\pi\Lambda^3 T$. 
Microscopically, the {\sl negative} ground state 
pressure is due to the (anti)caloron's 
holonomy shift by gluon exchange generating a constituent 
monopole and antimonopole 
\cite{Nahm1984,LeeLu1998,KraanVanBaalNPB1998,vanBaalKraalPLB1998,Brower1998} 
subject to a mutual force induced by quantum fluctuations 
\cite{Diakonov2004}. Notice that attraction is much more likely 
than repulsion \cite{Hofmann2005} explaining the {\sl negative} 
ground-state pressure that emerges after 
spatial coarse-graining. It is stressed 
that the potential $V(\phi)$ is unique: the usual 
shift ambiguity $V(\phi)\to V(\phi)+\mbox{const}$, 
which occurs due to the second-order 
nature of the Euler-Lagrange equations when applied to a 
given effective theory, is absent since the solution to these equation needs to be 
BPS saturated and periodic in the 
Euclidean time $\tau$ \cite{Hofmann2005,HerbstHofmann2004}. Notice the conceptual and technical 
differences to conventional approaches such as the hard-thermal-loop (HTL) effective theory \cite{HTL} which 
is a nonlocal theory for interacting soft and ultrasoft modes. While the HTL approach 
intergrates perturbative ultraviolet fluctuations into 
effective vertices it does not shed light on the stabilization of the 
infrared physics presumably associated with the magnetic sector of the theory. In contrast, the derivation of the 
phase $\frac{\phi}{|\phi|}$ implies that its existence owes to 
nonperturbative correlations residing in the magnetic sector. Upon a spatial 
coarse-graining (taking care of the UV physics) our emerging effective theory is local and evades the infrared 
problem by a dynamical gauge symmetry breaking generating massive modes. 

After an admissible rotation 
to unitary gauge, each of the two off-Cartan modes is seen to acquire a 
mass $m=2e|\phi|$ while the Cartan mode remains massless. Here $e$ denotes 
the effective gauge coupling after coarse-graining. As in \cite{HerbstHofmannRohrer2004} 
we will refer to off-Cartan modes as tree-level heavy (TLH) and to 
Cartan modes as tree-level massless (TLM). 
Demanding the invariance of Legendre transformations between thermodynamical 
quantities when going from the fundamental to the 
effective theory, a first-order evolution equation for 
$e=e(T)$ follows. There exists an attractor to the 
evolution. Namely, the behavior of $e(T)$ at low temperatures 
is independent of the initial value. The plateau, 
$e\equiv 8.89$, expresses the constancy of magnetic charge attributed to an 
isolated, screened monopole which is liberated by a dissociating, 
large-holonomy caloron \cite{Hofmann2005,Diakonov2004}. At 
$T_{c}=13.87\,\frac{\Lambda}{2\pi}$ the function $e$ diverges as 
$e(T)\sim -\log(T-T_{c})$. Thus for 
$T\searrow T_{c}$ a total screening of the isolated  
magnetic charge of a monopole takes place. This implies the latter's 
masslessness, its condensation\footnote{At $T_{c}$ the ground 
state becomes instable with respect to the formation of a large caloron 
holonomy \cite{Diakonov2004}.} and the decoupling of 
the TLH modes. 

Omitting in unitary-Coulomb 
gauge the `propagation' of the $0$-component of 
the TLM mode ($A_0^{a=3}$), it was shown in 
\cite{HerbstHofmannRohrer2004} that the two-loop 
correction is negative and and that the ratio of its modulus 
to the free quasiparticle pressure is 
at most $\sim 10^{-3}$. It peaks at $T\sim 3\,T_{c}$. 
Neglecting $\la A^3_0(x) A^3_0(y)\ra$ was justified by the 
observation that the real part of the electric screening 
mass $\lim_{\vec{p}\to 0}\sqrt{\Pi_{00}(p_0=0,\vec{p})}$ 
diverges. Here $\Pi_{\mu\nu}$ denotes the one-loop polarization tensor of 
the TLM mode \cite{Hofmann2005}. In the present work we investigate 
in detail how reliable such an approximation is. 

The paper is organized as follows: 
In Sec.\,\ref{Pre} we present the effective 
theory, list its Feynman rules in the 
physical gauge (unitary-Coulomb), and discuss 
the constraints on loop-momenta emerging 
from the spatial coarse-graining. A calculation of 
the polarization tensor for an on-shell TLM 
mode (associated with the photon when postulating 
that SU(2)$_{\tiny\mbox{CMB}}
\stackrel{\tiny\mbox{today}}=$U(1)$_Y$ \cite{Hofmann2005,Hofmann20051,GiacosaHofmann2005}) 
is performed in Sec.\,\ref{PoltensTLM}. We discuss the emergence of a 
gap in the low-frequency domain of the black-body spectrum at temperatures 
$\sim 2\cdots 4$\,$T_{\tiny\mbox{CMB}}$ where 
$T_{\tiny\mbox{CMB}}\sim 2.73\,$K. A table-top experiment is proposed accordingly. 
Moreover, we aim at an explanation for the stability of 
cold, dilute, old, and large atomic hydrogen clouds recently 
observed within our galaxy. In Sec.\,\ref{Pres2} 
we present improved estimates for subdominant 
two-loop pressure corrections. Next we re-calculate exactly the dominant 
correction involving the nonlocal diagram: the `propagation' of the $0$-component 
of the TLM mode is now taken in to account. 
On a qualitative level, our present results 
confirm those obtained in \cite{HerbstHofmannRohrer2004}. In the 
last section a summary and outlook on 
future research are presented.    

\section{Prerequisites\label{Pre}}

\begin{figure}
\begin{center}
\leavevmode
\leavevmode
\vspace{4.3cm}
\includegraphics{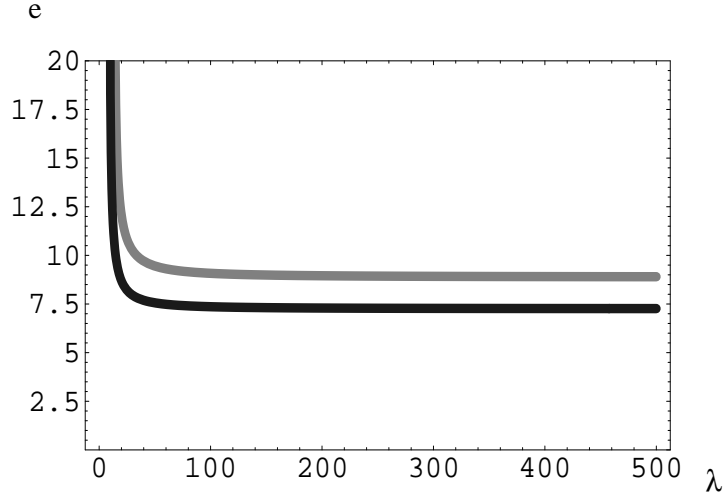}
\end{center}
\caption{\protect{\label{Fig-0A}} The one-loop evolution of the effective 
gauge coupling $e=e(\lambda)$ for SU(2) (grey curve) and SU(3) (black curve). 
The choice of the initial condition $a(\lambda_P=10^7)=0$ is arbitrary, the low-temperature 
behavior is unaffected by changing $\lambda_P$ as long as $\lambda_P$ is considerably larger 
than $\lambda_{c}$ \cite{Hofmann2005}.}      
\end{figure}

\begin{figure}
\begin{center}
\leavevmode
\leavevmode
\vspace{4.8cm}
\includegraphics{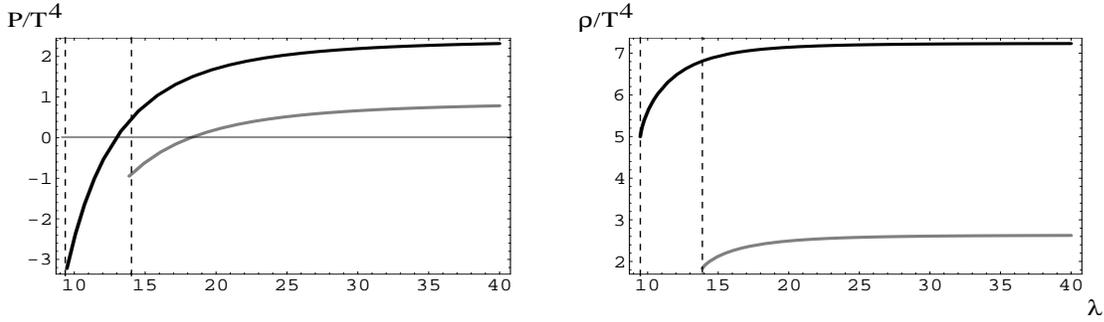}
\end{center}
\caption{\protect{\label{Fig-0B}}Ratio of the free quasiparticle pressure $P$ and $T^4$ (left panel) and 
the free quasiparticle energy density $\rho$ and $T^4$ (right panel) for SU(2) and SU(3) 
(grey and black, respectively) as a function of 
$\lambda\equiv\frac{2\pi T}{\Lambda}$ \cite{Hofmann2005}.}      
\end{figure}

Let us give a brief introduction into 
the effective theory for thermalized SU(2) Yang-Mills dynamics being in its 
deconfining phase \cite{Hofmann2005}. The following effective action emerges upon 
spatial coarse-graining down to a length scale equal to $|\phi|^{-1}=\sqrt{\frac{2\pi\,T}{\Lambda^3}}$:
\eqb
\label{act}
S = \mbox{tr}\, \int_0^\beta d\tau \int d^3x \left( \frac12\,G_{\mu\nu}G_{\mu\nu} +
D_\mu \phi D_\mu \phi + \Lambda^6 \phi^{-2} \right)\,.
\eqe
In Eq.\,(\ref{act}) $G_{\mu\nu}\equiv G^a_{\mu\nu} \frac{\lambda^a}{2}$, 
$G^a_{\mu\nu}=\pd_\mu A^a_\nu-\pd_\nu A^a_\mu+e\,\epsilon^{abc}A^b_\mu A^c_\nu$, and 
$D_\mu \phi=\partial_\mu \phi + ie[\phi,A_\mu]$ 
where $A_\mu$ is the (coarse-grained) gauge field of 
trivial topology, and $e$ denotes the effective gauge coupling. 
The latter can be extracted as
\eqb
\label{alam}
a=2\pi e\lambda^{-3/2}
\eqe
from the (inverted) 
solution of the (one-loop) evolution equation
\eqb
\label{evequ}
\pd_a\lambda=-\frac{24\lambda^4 a}{2\pi^6}\frac{D(2a)}{1+\frac{24\lambda^3 a^2}{2\pi^6}D(2a)}
\eqe
where 
\eqb
\label{Da}
D(a)\equiv \int_0^\infty dx\frac{x^2}{\sqrt{x^2+a^2}}\,\frac{1}{\exp(\sqrt{x^2+a^2})-1}\,,
\eqe
$\lambda\equiv\frac{2\pi T}{\Lambda}$, and $a\equiv \frac{m}{2T}$. The (coarse-grained) 
caloron-anticaloron ensemble is integrated into the 
nonfluctuating adjoint scalar field 
$\phi$ \cite{Hofmann2005,HerbstHofmann2004}. Eq.\,(\ref{evequ}) 
guarantees the invariance of the Legendre transformations between thermodynamical 
quantities when going from the fundamental to the effective theory. 
Notice that in deriving Eq.\,(\ref{evequ}) the vacuum part 
in the one-loop expression for the pressure $P$ 
can safely be neglected \cite{Hofmann2005}. In Figs.\,\ref{Fig-0A} and 
\ref{Fig-0B} the one-loop evolution of $e$ and that of 
$\frac{P}{T^4}$, $\frac{\rho}{T^4}$ are depicted.

Our calculations are performed within the real-time 
formulation of finite-tem\-pera\-ture field theory \cite{LandsmanWeert}. 
Let us formulate the Feynman rules:
The free propagator $D^{TLH,0}_{\mu\nu,ab}$ of a TLH mode in 
unitary-Coulomb gauge\footnote{This is a completely fixed, 
physical gauge in the effective theory after spatial coarse-graining: 
First one rotates $\phi$ and a pure-gauge ground-state field $a_\mu^{bg}$ 
to $\phi=\lambda_3|\phi|$ and $a_\mu^{bg}=0$. This is an admissible 
gauge rotation which, however, changes the value of the 
Polyakov loop (not taking the trace yet) ${\cal P}$ from $-\UM$ to $\UM$. Up to a finite 
renormalization this result coincides with the full expectation 
$\la {\cal P}\ra$ because there are massless TLM modes in 
the spectrum \cite{Hofmann2005}. Subsequently, one fixes the remaining U(1) 
gauge freedom by imposing $\pd_i A^{3}_i=0$. 
This gauge can always be reached with a periodic 
gauge function $\theta$ such that $\exp[i\theta]\in$U(1): 
$\theta(\tau=0,\vec{x})=\theta(\tau=\beta,\vec{x})$.} is
\begin{align}
\label{TLHprop}
D^{TLH,0}_{\mu\nu,ab}(p)&= -\delta_{ab}\tilde{D}_{\mu\nu}
\left[\frac{i}{p^2-m^2}+2\pi\delta(p^2-m^2)n_B(|p_0|/T) \right]\\
\tilde{D}_{\mu\nu} &= \left( g_{\mu\nu}-\frac{p_\mu p_\nu}{m^2} \right)
\end{align}
where $n_B(x)=1/(e^x-1)$ denotes the Bose-Einstein distribution function. 
For the free TLM mode we have
\begin{equation}
\label{TLMprop}
D^{TLM,0}_{ab,\mu\nu}(p) = -\delta_{ab}
\left\lbrace P^T_{\mu\nu}\left[\frac{i}{p^2}+2\pi\delta(p^2)n_B(|p_0|/T)\right]
-i\frac{u_\mu u_\nu}{\textbf{p}^2}\right\rbrace \,.
\end{equation} 
where 
\begin{align}
P^{00}_T & = P^{0i}_T = P^{i0}_T = 0\\
P^{ij}_T & = \delta^{ij} - p^{i}p^{j}/\textbf{p}^2\,.
\end{align}
TLM modes carry a color index 3 while TLH 
modes have a color index 1 and 2. Notice the term $\propto u_\mu u_\nu$ in 
Eq.\,(\ref{TLMprop}) describing the 'propagation' of the 
$A^3_0$ field. Here $u_\mu=(1,0,0,0)$ represents the four-velocity of the heat bath.   

\begin{figure}
\begin{center}
\leavevmode
\leavevmode
\vspace{4.9cm}
\includegraphics{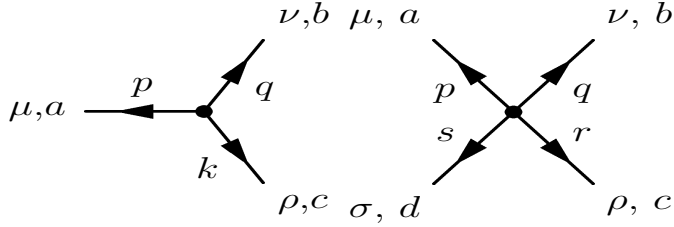}
\end{center}
\caption{\protect{\label{Fig-34vert}} Three- and four-vertices.}      
\end{figure}

\noindent The three- and four-gauge-boson vertices are given as
\begin{align}
\Gamma^{\mu\nu\rho}_{[3]abc}(p,k,q) &= e (2\pi)^4\delta(p+q+k) f_{abc} [g^{\mu\nu}(q-p)^\rho + g^{\nu\rho}(k-q)^\mu + g^{\rho\mu}(p-k)^\nu]\\
\begin{split}
\Gamma^{\mu\nu\rho\sigma}_{[4]abcd} &= -ie^2(2\pi)^4\delta(p+q+s+r)[f_{abe}f_{cde}(g^{\mu\rho}g^{\nu\sigma}-g^{\mu\sigma}g^{\nu\rho})\\
& \quad +f_{ace}f_{bde}(g^{\mu\nu}g^{\rho\sigma}-g^{\mu\sigma}g^{\nu\rho})\\
& \quad +f_{ade}f_{bce}(g^{\mu\nu}g^{\rho\sigma}-g^{\mu\rho}g^{\nu\sigma})]
\end{split}
\end{align}
with the four momenta and color indices defined in Fig.\,\ref{Fig-34vert}.
According to \cite{LandsmanWeert} one has to divide a loop-diagram by $i$ and by 
the number of its vertices.

The analytical description of the nontrivial ground-state dynamics is 
facilitated by a spatial coarse-graining down to a resolution $|\phi|$. 
Thus the maximal off-shellness of gauge modes is constrained\footnote{The idea of a Wilsonian renormalization-group
flow is realized after the interacting topologically nontrivial sector has been integrated 
out. At each loop order (expansion in $\hbar^{-1}$) a modification of the on-shell condition 
emerges which determines the restriction on admissible off-shellness for the next loop order.} 
to be \cite{Hofmann2005} 
\begin{equation}\label{constraint1}
|p^2-m^2|\leq|\phi|^2
\end{equation}
where $m=0$ for a TLM mode and $m=2e\,|\phi|=2e\sqrt{\frac{\Lambda^3}{2\pi T}}$ for 
a TLH mode. Moreover, the resolution associated with 
a four vertex of ingoing momenta $p,k$ is constrained as 
\eqb
\label{constraint2}
|(p+k)^2|\leq|\phi|^2\,.
\eqe
Notice that the constraint in (\ref{constraint2}) is only applicable if two of the four legs
associated with the vertex form a closed 
loop. If this is not the case then one needs to distinguish $s$-, $t$-, 
and $u$-channel scattering.
 
\section{One-loop polarization tensor for an on-shell TLM mode \label{PoltensTLM}}

\subsection{General considerations}

Here we compute the diagonal components of the polarization 
tensor $\Pi^{\mu\nu}(p_0,\vec{p})$ for the TLM mode at the one-loop level specializing 
to $p^2=0$. This is conceptually interesting and provides for a
prediction associated with a modification of the low-momentum part 
of black body spectra at low temperatures. Moreover, a computation of 
$\Pi^{ii}(p_0,\vec{p})$ is necessary for an analytical grasp of the physics related to 
the large-angle regime in CMB maps. 

For any value of $p^2$ the polarization tensor $\Pi^{\mu\nu}$ for the TLM mode is 
transversal,
\eqb
\label{transPi}
p_\mu \Pi^{\mu\nu}=0\,.
\eqe
Hence the following decomposition holds
\eqb
\label{Pidec}
\Pi^{\mu\nu}=G(p_0,\vec{p})\,P^{\mu\nu}_T+F(p_0,\vec{p})\,P^{\mu\nu}_L
\eqe
where 
\eqb
\label{PL}
P^{\mu\nu}_L\equiv \frac{p^\mu p^\nu}{p^2}-g^{\mu\nu}-P^{\mu\nu}_T\,.
\eqe
The functions $G(p_0,\vec{p})$ and $F(p_0,\vec{p})$ determine the propagation of 
the interacting TLM mode. For $\mu=\nu=0$ Eq.\,(\ref{Pidec}) yields upon rotation 
to real-time 
\eqb
\label{P00F}
F(p_0,\vec{p})=\left(1+\frac{p_0^2}{p^2}\right)^{-1}\,\Pi^{00}\,.
\eqe
Assuming $\vec{p}$ to be parallel to the $z$-axis, we have
\eqb
\label{gxxgyy}
\Pi_{11}=\Pi_{22}=G(p_0,\vec{p})\,.
\eqe
In the Euclidean formulation the interacting 
propagator $D^{TLM}_{ab,\mu\nu}(p)$ reads
\eqb
\label{propTLMWW}
D^{TLM}_{ab,\mu\nu}(p) = -\delta_{ab}
\Big \lbrace P^T_{\mu\nu}\frac{1}{G-p^2}+\frac{p^2}{\vec{p}^2}\frac{1}{F-p^2}\,u_\mu u_\nu\Big \rbrace\,.
\eqe
Notice that for $F=G=0$ and rotating to Minkowskian signature 
Eq.\,(\ref{propTLMWW}) transforms into Eq.\,(\ref{TLMprop}). 
In \cite{Hofmann2005} $\Pi^{00}$ was calculated for $p_0=0$ 
and in the limit $\vec{p}\to 0$. One has 
$|\Pi^{00}(0,\vec{p}\to 0)|=|F(0,\vec{p}\to 0)|=\infty$. 
According to Eq.\,(\ref{propTLMWW}) the term $\propto u_\mu u_\nu$ vanishes in this limit. 
One of the tasks of the present paper is to check how reliable it is to neglect this term altogether when calculating the two-loop corrections to the pressure.

Going on-shell, $|p_0|\to |\vec{p}|$, in Eq.\,(\ref{P00F}), 
we observe that $F\to 0$ provided that $\Pi^{00}$ remains 
finite in this limit. We have computed $\Pi^{00}$ for $p^2=0$ 
and we have seen that this is, indeed, the case. 
This, in turn, implies 
that the longitudinal structure (quantum `propagation') in Eq.\,(\ref{propTLMWW}) reduces 
to the free limit as in Eq.\,(\ref{TLMprop}). 
The function $G$ modifies the dispersion law for the TLM mode 
as follows:
\eab
\label{displaw}
\omega^2(\vec{p})&=&\vec{p}^2+\mbox{Re}\,G(\omega(\vec{p}),\vec{p})\,,\nonumber\\ 
\gamma(\vec{p})&=&-\frac{1}{2\omega}\mbox{Im}\,G(\omega(\vec{p}),\vec{p})\,.
\eae
For technical simplicity we evaluate $\Pi_{11}=G$ for $p_0=|\vec{p}|$. 
Assuming that $G(\omega(\vec{p}),\vec{p})$ is analytic about 
$\omega=|\vec{p}|$ and that it depends only weakly on $\omega$, 
an interpretation of this result in the sense of Eq.\,(\ref{displaw}) is 
facilitated. Namely, setting $\omega=|\vec{p}|$, the right-hand 
side of Eq.\,(\ref{displaw}) represents a useful approximation to the exact solution 
$\omega^2(\vec{p})$ which is expanded in powers of $\omega-|\vec{p}|$. Notice that
the constraint in (\ref{constraint1}) does not allow for a large deviation 
from $\omega=|\vec{p}|$ anyhow. 

\subsection{Calculation of $G$\label{CalcPi}}
  
\begin{figure}
\begin{center}
\leavevmode
\leavevmode
\vspace{4.9cm}
\includegraphics{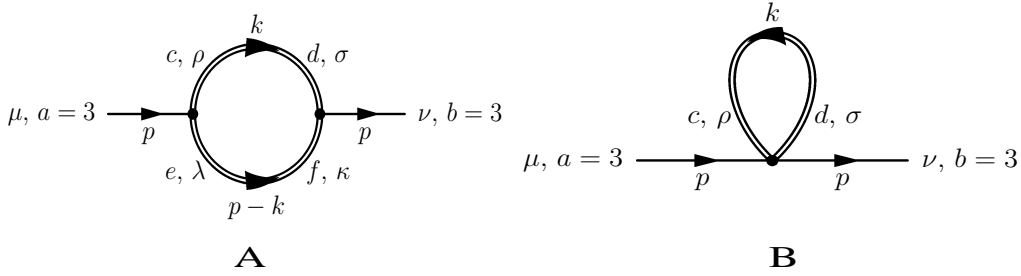}
\end{center}
\caption{\protect{\label{Fig-nll}} The diagrams for the TLM mode polarization tensor.}      
\end{figure}

$\Pi^{\mu\nu}$ is the sum of the 
two diagrams A and B in Fig.\,\ref{Fig-nll}. 
For $p^2=0$ diagram A vanishes. This can be seen as follows. We have 
\begin{equation}
\label{Anproc}
\begin{split}
\Pi^{\mu\nu}_{A}(p)\,=\,& 
\frac{1}{2i}\int\frac{d^4k}{(2\pi)^4} e^2 
	\epsilon_{ace}[g^{\mu\rho}(-p-k)^\lambda+g^{\rho\lambda}(k-p+k)^\mu+g^{\lambda\mu}(p-k+p)^\rho]\times\\
& \epsilon_{dbf}[g^{\sigma\nu}(-k-p)^\kappa+g^{\nu\kappa}(p+p-k)^\sigma+g^{\kappa\sigma}(-p+k+k)^\nu]\times\\
&(-\delta_{cd})\left(g_{\rho\sigma}-\frac{k_\rho k_\sigma}{m^2}\right) 
\left[\frac{i}{k^2-m^2}+2\pi\delta(k^2-m^2)\,n_B(|k_0|/T) \right]\times\\
&(-\delta_{ef})\left(g_{\lambda\kappa}-\frac{(p-k)_\lambda(p-k)_\kappa}{(p-k)^2}\right)\times\\
&\left[\frac{i}{(p-k)^2-m^2}+2\pi\delta((p-k)^2-m^2)\,n_B(|p_0-k_0|/T) \right]
\end{split}
\end{equation}
From the one-loop evolution \cite{Hofmann2005} we know that $e\ge 8.89$. Due to constraint 
(\ref{constraint1}) the vacuum part in the TLH propagator thus is 
forbidden. Using $p^2=0$ and $k^2=m^2$, the thermal part 
$\Pi^{\mu\nu}_{A,\tiny\mbox{therm}}(p)$ reads
\begin{equation}\label{pitherm}
\begin{split}
\Pi^{\mu\nu}_{A,\tiny\mbox{therm}}(p)\,=\,&
ie^2\int\frac{d^4k}{(2\pi)^2} \Big[ \left(2kp-4\frac{(kp)^2}{m^2}\right)g^{\mu\nu} +  
\left(12-2\frac{kp}{m^2}\right)k^\mu k^\nu+\\
& \left(-6+4\frac{kp}{m^2}\right)(k^\nu p^\mu+k^\mu p^\nu)
 +\left(-5+\frac{(kp)^2}{m^4}\right)p^\mu p^\nu \Big]\times\\
& \delta(k^2-m^2)\,n_B(|k_0|/T)\,\delta((p-k)^2-m^2)\,n_B(|p_0-k_0|/T)\,.
\end{split}
\end{equation}
For $p_0>0$ the product of $\delta$-functions can be rewritten as
\begin{equation}
\label{PAdeltas}
\begin{split}
\delta(k^2-m^2)\cdot\delta((p-k)^2-m^2)\,=\,&  
\frac{1}{4 p_0 |\textbf{k}| \sqrt{|\textbf{k}|^2+m^2}}\times\\ 
&\Big[\delta\left(k_0-\sqrt{|\textbf{k}|^2+m^2}\right)
\cdot\delta\left(\cos\theta-\frac{\sqrt{|\textbf{k}|^2+m^2}}{|\textbf{k}|}\right)+\\
&\delta\left(k_0+\sqrt{|\textbf{k}|^2+m^2}\right) \cdot\delta\left(\cos\theta+
 \frac{\sqrt{|\textbf{k}|^2+m^2}}{|\textbf{k}|}\right)\Big]\,
\end{split}
\end{equation}
where $\theta\equiv\angle(\vec{p},\vec{k})$. Because 
$\frac{\sqrt{|\textbf{k}|^2+m^2}}{|\textbf{k}|}>1$ and $-1\leq\cos\theta\leq1$ 
the argument of the two $\delta$-functions in Eq.\,(\ref{PAdeltas})  
never vanishes and thus the right-hand of Eq.(\ref{pitherm}) is zero.

\noindent Applying the Feynman rules of Sec.\,\ref{Pre}, diagram B 
reads
\begin{equation}
\label{vactad}
\begin{split}
\Pi^{\mu\nu}_{B}(p)\,=\,&\frac{1}{i} \int \frac{d^4k}{(2\pi)^4} 
(-\delta_{ab}) \left( g_{\rho\sigma}-\frac{k_\rho k_\sigma}{m^2} \right)  
\left[\frac{i}{k^2-m^2}+2\pi\delta(k^2-m^2)n_B(|k_0|/T) \right]\times\\
 & \quad (-ie^2)[
 \epsilon_{abe}\epsilon_{cde}(g^{\mu\rho}g^{\nu\sigma}-g^{\mu\sigma}g^{\nu\rho})
 +\epsilon_{ace}\epsilon_{bde}(g^{\mu\nu}g^{\rho\sigma}-g^{\mu\sigma}g^{\nu\rho})+\\
 &\quad\epsilon_{ade}\epsilon_{bce}(g^{\mu\nu}g^{\rho\sigma}-g^{\mu\rho}g^{\nu\sigma})]\,.
\end{split}
\end{equation}
Again, the part in Eq.\,(\ref{vactad}) arising from the vacuum contribution in 
Eq.\,(\ref{TLHprop}) vanishes because of constraint (\ref{constraint1}). 
Applying constraint (\ref{constraint2}) for $p_0>0,\,p^2=0$ yields:
\begin{equation}
\label{condproc}
\begin{split}
|(p+k)^2|\,=\,&|2p\,k+k^2| = \left|2p_0(k_0-|\textbf{k}|\cos\theta)+4e^2|\phi|^2\right|\\
\,=\,&\left|2 p_0\left(\pm\sqrt{\textbf{k}^2+4e^2|\phi|^2}-|\textbf{k}|\cos\theta\right) + 4e^2|\phi|^2\right|
 {\leq} |\phi|^2\,.
\end{split}
\end{equation}
For the $+$ sign the condition in Eq.\,(\ref{condproc}) is never satisfied, for the $-$ sign there is a 
range for $p_0$ where the condition is not violated. 

We are interested in $\frac{\Pi_{11}}{T^2}=\frac{\Pi_{22}}{T^2}=\frac{G}{T^2}$ as a function of
$X\equiv\frac{|\vec{p}|}{T}$ and $\lambda\equiv\frac{2\pi T}{\Lambda}$ when $\vec{p}$ is 
parallel to the $z$-axis. Performing 
the $k_0$-integration in Eq.\,(\ref{vactad}) and introducing dimensionless variables as
\eqb
\label{vardimless}
\vec{y}\equiv\frac{\vec{k}}{|\phi|}\,,
\eqe
we obtain from Eq.\,(\ref{vactad}) that
\eab
\label{dimlPixx}
\frac{G}{T^2}&=&\frac{\Pi_{11}}{T^2}=\frac{\Pi_{22}}{T^2}=\nonumber\\ 
&=&\frac{e^2}{\pi\lambda^3}\,\int
d^3y\,\left(-2+\frac{y_1^2}{4e^2}\right)\,
\frac{n_B\left(2\pi\lambda^{-3/2}\sqrt{\vec{y}^2+4e^2}\right)}{\sqrt{\vec{y}^2+4e^2}}
\eae
where the integration is subject to the following constraint:
\eqb
\label{constdimless}
-1\le -\lambda^{3/2}\,\frac{X}{\pi}\,\left(\sqrt{\vec{y}^2+4e^2}+y_3\right)+4e^2\le 1\,.
\eqe
In view of constraint (\ref{constdimless}) the integral in Eq.\,(\ref{dimlPixx}) is 
evaluated most conveniently in cylindrical coordinates,
\eqb
\label{cylcoo}
y_1=\rho\,\cos\varphi\,,\ \ \ \ y_2=\rho\,\sin\varphi\,,\ \ \ \ y_3=\xi\,.
\eqe
Let us now discuss how the constraint (\ref{constdimless}) is implemented 
in the $\rho$- and $\xi$-integration. Constraint (\ref{constdimless}) is re-cast as
\eqb
\label{constrec}
\frac{4e^2-1}{\lambda^{3/2}}\,\frac{\pi}{X}\le \sqrt{\rho^2+\xi^2+4e^2}+
\xi\le\frac{4e^2+1}{\lambda^{3/2}}\,\frac{\pi}{X}\,.
\eqe
Notice that Eq.\,(\ref{constrec}) gives an upper bound $\Xi$ for $\xi$: 
$\xi<\frac{4e^2+1}{\lambda^{3/2}}\,\frac{\pi}{X}\equiv\Xi$. In contrast, there is no such 
global lower bound for $\xi$.  

The upper limits for the $\rho$- and $\xi$-integration 
are obtained as follows. Since $\xi<\Xi$ we can square the second part of the inequality 
(\ref{constrec}) and solve for $\rho$:
\eqb
\label{rhoM}
\rho\le\sqrt{\left(\frac{\pi}{X}\right)^2\,\frac{(4e^2+1)^2}{\lambda^3}-
\frac{2\pi}{X}\,\frac{4e^2+1}{\lambda^{3/2}}\xi-4e^2}\equiv \rho_M(X,\xi,\lambda)\,.
\eqe
The condition that the expression under the square root in 
Eq.\,(\ref{rhoM}) is positive yields the upper limit $\xi_M(X,\lambda)$ for the 
$\xi$-integration:
\eqb
\label{xiM}
\xi\le \frac{\pi}{2X}\,\frac{4e^2+1}{\lambda^{3/2}}-2\,\frac{X}{\pi}\,\lambda^{3/2}\,
\frac{e^2}{4e^2+1}\equiv\xi_M(X,\lambda)\,.
\eqe
\noindent The lower limit for the $\rho$-integration 
is obtained as follows. Upon subtracting $\xi$ from the first 
part of the inequality (\ref{constrec}) the result can be 
squared provided that $\xi<\frac{4e^2-1}{\lambda^{3/2}}\,\frac{\pi}{X}$. Solving for $\rho$, 
we have
\eqb
\label{rholow}
\rho\ge\sqrt{\left(\frac{\pi}{X}\right)^2\,\frac{(4e^2-1)^2}{\lambda^3}-
\frac{2\pi}{X}\,\frac{4e^2-1}{\lambda^{3/2}}\xi-4e^2}\equiv \rho_m(X,\xi,\lambda)\,.
\eqe
The condition that the expression under the square root in 
Eq.\,(\ref{rholow}) is positive introduces the critical value $\xi_m(X,\lambda)$ for the 
$\xi$-integration as:
\eqb
\label{xim}
\xi_m(X,\lambda)\equiv \frac{\pi}{2X}\,\frac{4e^2-1}{\lambda^{3/2}}-2\,\frac{X}{\pi}\,\lambda^{3/2}\,
\frac{e^2}{4e^2-1}\,.
\eqe
For $-\infty<\xi\le\xi_m(X,\lambda)$ the lower limit for the $\rho$-integration is given by 
$\rho_m(x,\xi,\lambda)$. Notice that according to Eq.\,(\ref{xim}) $\xi_m(X,\lambda)$ is always smaller than 
$\frac{4e^2-1}{\lambda^{3/2}}\,\frac{\pi}{X}$ such that our 
above assumption is consistent. According to Eq.\,(\ref{constrec}), the 
opposite case, $\frac{4e^2-1}{\lambda^{3/2}}\le\xi\le\xi_M(X,\lambda)$, leads to $\rho\ge 0$ 
which does not represent an additional constraint. To summarize, we have
\eab
\label{intlimts}
\frac{G}{T^2}&=&\left[\int_{-\infty}^{\xi_m(X,\lambda)}d\xi\,
\int_{\rho_m(X,\xi,\lambda)}^{\rho_M(X,\xi,\lambda)} d\rho+
\int_{\xi_m(X,\lambda)}^{\xi_M(X,\lambda)} d\xi\,\int_0^{\rho_M(X,\xi,\lambda)} d\rho\right]\nonumber\\ 
&&e^2\lambda^{-3}\left(-4+\frac{\rho^2}{4e^2}\right)\,\rho\,\frac{n_B\left(2\pi \lambda^{-3/2}\sqrt{\rho^2+\xi^2+4e^2}\right)}
{\sqrt{\rho^2+\xi^2+4e^2}}\,
\eae
where the integral operation indicated in the square brackets is applied to the last line.

\subsection{Results and discussion} 

\subsubsection{General discussion of results\label{GDR}}

\begin{figure}
\begin{center}
\leavevmode
\leavevmode
\vspace{4.3cm}
\includegraphics{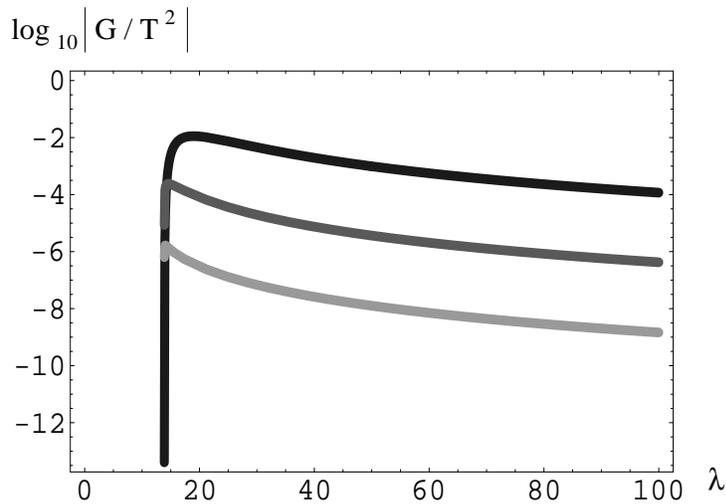}
\end{center}
\caption{\protect{\label{Fig-4}}$\left|\frac{G}{T^2}\right|$ as a function of 
$\lambda\equiv\frac{2\pi T}{\Lambda}$ for $X=1$ (black), $X=5$ (dark grey), and $X=10$ (light grey).}      
\end{figure}

\begin{figure}
\begin{center}
\leavevmode
\leavevmode
\vspace{5.4cm}
\includegraphics{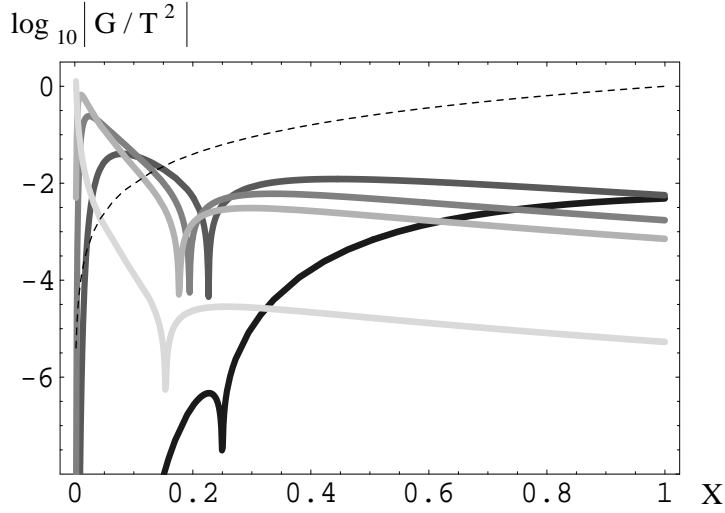}
\end{center}
\caption{\protect{\label{Fig-5}}$\left|\frac{G}{T^2}\right|$ as a function of $X\equiv\frac{|\vec{p}|}{T}$ 
for $\lambda=1.12\,\lambda_{c}$ (black), $\lambda=2\,\lambda_{c}$ (dark grey), 
$\lambda=3\,\lambda_{c}$ (grey), $\lambda=4\,\lambda_{c}$ (light grey), $\lambda=20\,\lambda_{c}$ 
(very light grey). The dashed curve is a 
plot of the function $f(X)=2\log_{10}X$. TLM modes are strongly screened at 
$X$-values for which $\log_{10}\left|\frac{G}{T^2}\right|>f(X)$ ($\frac{\sqrt{G}}{T}>X$), 
that is, to the left of the dashed line.}      
\end{figure}

In Figs.\,\ref{Fig-4}, \ref{Fig-5}, and \ref{Fig-lowmom} we show plots 
of $\log_{10}\left|\frac{G}{T^2}\right|$ as obtained 
by a numerical integration. We have used the one-loop evolution of the 
effective coupling $e=e(\lambda)$. For all $\lambda$ and values of $X$ to the right of the 
dips in Fig.\,\ref{Fig-5} $\frac{G}{T^2}$ is negative and real (antiscreening). According to the dispersion law in 
Eq.\,(\ref{displaw}) this implies that the energy of a propagating TLM mode 
is {\sl reduced} as compared to the free
case. For $X$ values to the left of the dips $\frac{G}{T^2}$ 
is positive and real (screening) with interesting consequences for the low-momentum regime 
of black-body spectra, see Sec.\,\ref{CMB}.    
Fig.\,\ref{Fig-4} indicates the dependence of 
$\log_{10}\left|\frac{G}{T^2}\right|$ on $\lambda$ 
keeping $X=1,5,10$ fixed. Obviously, the effect on the propagation of TLM modes arising 
from TLH intermediate states is very small 
(maximum of $\left|\frac{G}{T^2}\right|$ at $X=1$: $\sim 10^{-2}$). As for the high-temperature 
behavior we observe the following. On the one hand, there is clear evidence for a power-like 
suppression of $\left|\frac{G}{T^2}\right|$ in $\lambda$. Recall that the 
one-loop result for the (quasiparticle) pressure shows a power-like approach 
to the Stefan-Boltzmann limit \cite{Hofmann2005}. In a similar way, the approach to the limit of vanishing
antiscreening is also power-like for the TLM mode. On the one hand, a sudden drop of $\left|\frac{G}{T^2}\right|$ 
occurs for $\lambda\searrow \lambda_{c}=13.87$. This signals that intermediary TLH modes decouple due to 
their diverging mass. On the other hand, the values of 
$\log_{10}\left|\frac{G}{T^2}\right|$ at fixed 
$\lambda>\lambda_{c}$ are equidistant for (nearly) 
equidistant values of $X$. This shows the exponential suppression of 
$\log_{10}\left|\frac{G}{T^2}\right|$ for $X\ge 1$ and can be 
understood as follows. For large $X$ Eq.\,(\ref{constrec}) demands $\xi$ to be negative 
and $|\xi|,\rho$ to be large. As a consequence, the square root in 
Eq.\,(\ref{constrec}), which appears as an argument of $n_B$, see Eq.\,(\ref{intlimts}), 
is large thus implying exponential suppression in $X$.

Fig.\,\ref{Fig-5} indicates the dependence of 
$\log_{10}\left|\frac{G}{T^2}\right|$ on $X$ 
keeping $\lambda=1.12\lambda_{c}$, $\lambda=2\lambda_{c}$, $3\lambda_{c}$, 
$4\lambda_{c}$, and  $\lambda=20\lambda_{c}$ fixed. Notice that the low-momentum regime is 
investigated. Namely, for $X\sim 1$ the afore-mentioned exponential suppression 
sets in which can be seen by the linear decrease. For $X<0.6$ the black curve 
($\lambda=1.12\lambda_{c}$) is below the curves for $\lambda=2\lambda_{c}, 3\lambda_{c}, 
4\lambda_{c}$ because of the vicinity of $\lambda$ to $\lambda_{c}$ where the mass 
of TLH modes diverges. The smallness of the curve for $\lambda=20\lambda_{c}$ arises due 
to the above-discussed power suppression. Notice the sharp dip 
occurring for $X$-values in the 
range $0.15\le X\le 0.25$. The dip is caused by a change in sign for 
$G$: For $X$ to the right of the dip $G$ is negative (antiscreening) 
while it is positive (screening) to the left. The dashed 
line is a plot of the function $f(X)=2\log_{10}X$. The intersection 
of a curve with $f(X)$ indicates the momentum where the dynamical mass of the 
TLM mode is equal to the modulus of its spatial momentum (strong screening), 
see Eq.\,(\ref{displaw}). For $\lambda\sim \lambda_{c}$ or for 
$\lambda\gg\lambda_{c}$ the strong-screening regime shrinks to the point 
$X=0$. While $\left|\frac{G}{T^2}\right|$ is practically zero in the 
former case it is sizable in the latter. For $\lambda=2\lambda_{c}, 3\lambda_{c}, 4\lambda_{c}$ 
Fig.\,\ref{Fig-5} shows that the strong-screening regime has a 
finite support beginning at $X_s\sim 0.15$. The fact that within our approximation 
$p^2=0$ no imaginary part is being generated in 
$\frac{G}{T^2}$ is related to the vanishing of diagram A, 
see Eq.\,(\ref{Anproc}). For $p^2\not=0$ diagram A 
is purely imaginary and thus would lead to an additional 
damping in propagating modes. Since we are interested in TLM 
modes $p^2\sim 0$ we expect the effect of 
diagram A to be negligible.

\subsubsection{Application to SU(2)$_{\tiny\mbox{CMB}}$\label{CMB}}

In \cite{Hofmann2005,Hofmann20051} the postulate was put forward that 
the photon (regarded as a propagating wave) is identified with the 
TLM mode of an SU(2) Yang-Mills theory with scale 
$\Lambda\sim 10^{-4}\,$eV: SU(2)$_{\tiny\mbox{CMB}}$. For the present 
cosmological epoch we have $T_{c}=T_{\tiny\mbox{CMB}}\sim 2.73\,$K$\sim 2.35\times 10^{-4}\,$eV. 
The viability of such a postulate was discussed and 
checked in view of cosmological and astrophysical bounds in 
\cite{GiacosaHofmann2005}. 

As discussed in Sec.\,\ref{GDR} today's photon, which propagates above the 
ground state of SU(2)$_{\tiny\mbox{CMB}}$, is unaffected 
by nonabelian fluctuations (TLH modes) because of the 
decoupling of the latter. For radiation of a 
temperature considerably above $T_{\tiny\mbox{CMB}}$, for example room temperature ($\sim \frac{1}{40}\,$eV), 
a detection of the distortion of the black-body spectrum at low momenta 
surely is outside the reach of 
experiments, see Fig.\,\ref{Fig-5}. This is, however, no longer 
true if the temperature is a few times $T_{\tiny\mbox{CMB}}$. 

We thus propose the following table-top experiment for an independent check of 
the above postulate: The wave-length $l_s$ 
of the strongly screened mode ($X<X_s=0.15$) at $T=2\,T_{\tiny\mbox{CMB}}$ (about the 
boiling temperature of liquid $^4$He) is given as 
\eqb
\label{critwl}
l_{s}=\frac{h\,c}{k_B\,T\,X_s}=1.8\,\mbox{cm}\,.
\eqe
To detect the absence of low-momentum modes ($X_s=0.15$) in the black-body spectrum 
at $T=2\,T_{\tiny\mbox{CMB}}$, at least one linear dimension $d$ of the isolated 
cavity should be considerably larger than 1.8\,cm, say $d\sim\,$50\,cm. 

In \cite{Morozova1993} the construction of a low-temperature black body 
(LTBB), to be used as a temperature normal, was reported for temperature ranges 80\,K$\le T\le 300$\,K. For the 
lower limit $T=80\,$K we obtain $X_s=0.0366$ corresponding to 
$l_s=0.49\,$cm. 

Let us now discuss how sensitive the measurement of the LTBB spectral intensity $I(X)$ needs to 
be in order to detect the spectral gap setting in at $X_s$. A useful criterion 
is determined by the ratio $R(X_s)$ of $I(X_s)$ and $I(X_{\tiny\mbox{max}})$ where 
$X_{\tiny\mbox{max}}=2.82$ is the position 
of the maximum of $I(X)$ (back to natural units):
\eqb
\label{R(xt)}
R(X_s)\equiv\frac{I(X_s)}{I(X_{\tiny\mbox{max}})}=\frac{1}{1.42144}\,\,\frac{X_s^3}{\exp(X_s)-1}\,.
\eqe
For $T=80\,$K we have $R(X_s=0.0366)=9\times 10^{-4}$. To achieve such a high precision is a 
challenging task. To the best of the authors knowledge in 
\cite{Morozova1993} only the overall and not the spectral 
intensity of the LTBB was measured. For $T=5\,$K one has $R(X_s=0.14)=1.2\times 10^{-2}$. 
Thus at very low temperatures the precision required to detect the spectral gap is within the 
1\%-range. It is, however, experimentally challenging to cool 
the LTBB down to these low temperatures. To the best of the authors 
knowledge a precision measurement of the low-frequency regime of a LTBB 
at $T=5\cdots 10\,$K has not yet been performed. We believe 
that such an experiment is well feasible: It will represent an important and inexpensive 
test of the postulate SU(2)$_{\tiny\mbox{CMB}}\stackrel{\tiny\mbox{today}}=$U(1)$_Y$.

\subsubsection{Possible explanation of the stability of innergalactic 
clouds of atomic hydrogen at $T\sim 5\cdots 10\,$K\label{cH1}}

In \cite{H1KB,Dickey2001} the existence of a large (up to 2\,kpc), old (estimated age $\sim$ 50 million years), 
cold (mean brightness temperature $T_B\sim 20\,$K with cold regions of $T_B\sim 5\cdots 10\,$K), 
dilute (number density: $\sim 1.5$\,cm$^{-3}$) and massive ($1.9\times 10^7$ solar masses) 
innergalactic cloud (GSH139-03-69) of atomic hydrogen (HI) forming an arc-like structure in between spiral arms 
was reported. In \cite{Dickey} and references therein smaller structures of this type were identified. 
These are puzzling results which do not fit into the dominant model 
for the interstellar medium \cite{Dickey2001}. Moreover, considering the typical 
time scale for the formation of H$_2$ out of HI of about $10^{6}\,$yr 
\cite{Dickey} at these low temperatures and low densities clashes with the inferred age 
of the structure observed in \cite{H1KB}.

\begin{figure}
\begin{center}
\leavevmode
\leavevmode
\vspace{4.9cm}
\includegraphics{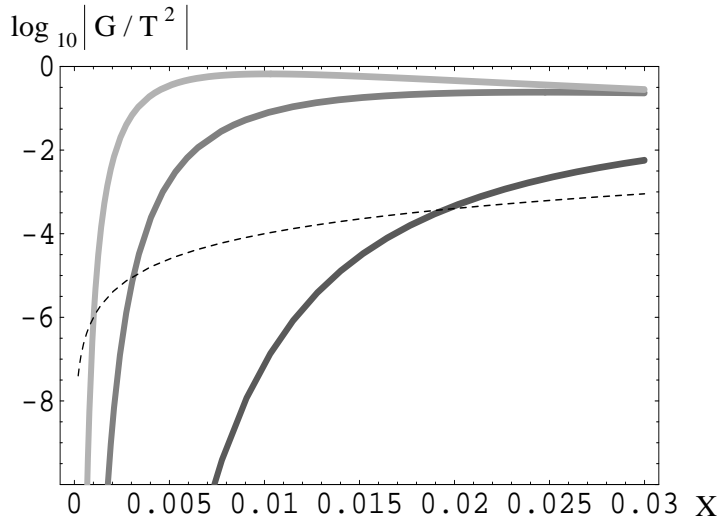}
\end{center}
\caption{\protect{\label{Fig-lowmom}}$\left|\frac{G}{T^2}\right|$ as a function of $X\equiv\frac{|\vec{p}|}{T}$ 
for $\lambda=2\,\lambda_{c}$ (dark grey), 
$\lambda=3\,\lambda_{c}$ (grey), and $\lambda=4\,\lambda_{c}$ (light grey) and very low momenta. 
The dashed curve is a 
plot of the function $f(X)=2\log_{10}X$. TLM modes are strongly screened for 
$X>X_{\tiny\mbox{low}}$ for which $\log_{10}\left|\frac{G}{T^2}\right|>f(X)$.}      
\end{figure}

To the best of the authors knowledge there is no standard 
explanation for the existence and the stability 
of such structures. We wish to propose a scenario possibly 
explaining the stability based on SU(2)$_{\tiny\mbox{CMB}}$. 
Namely, at temperatures $T_B\sim 5\cdots 10\,$K, corresponding to $T_B\sim 2\cdots 4\,T_{\tiny\mbox{CMB}}$, 
the polarization tensor of photons with momenta ranging 
between $|\vec{p}_s|=0.15\,T_B>|\vec{p}_c|>|\vec{p}_{\tiny\mbox{low}}|$ is such 
that it forbids their propagation (strong screening), see Figs.\,\ref{Fig-5} and 
\ref{Fig-lowmom}, where $|\vec{p}_{\tiny\mbox{low}}|$ depends rather 
strongly on temperature (Fig.\,\ref{Fig-lowmom}).     

Incidentally, the regime for the wavelength $l_c$ associated with $|\vec{p}_c|$ is 
comparable to the interatomic distance $\sim 1\,$cm in GSH139-03-69: 
At $T=5\,$K we have $l_s=2.1\,\mbox{cm}\le l_c\le 8.8\,\mbox{cm}=l_{\tiny\mbox{low}}$, at 
$T=10\,$K we have $l_s=1.2\,\mbox{cm}\le l_c\le 1.01\,\mbox{m}=l_{\tiny\mbox{low}}$. Thus the 
(almost on-shell) photons being emitted by a given HI particle to mediate the dipole interaction towards another 
HI particle are far off their mass shell at typical interatomic distances. As a consequence, 
the dipole force at these distances appears 
to be switched off: H$_2$ molecules are prevented from 
forming at these temperatures and densities.  

At this point we would like to discuss an apparent paradox involving the concept 
of a spin temperature $T_S$ for GSH139-03-69. The latter is defined as a 
temperature associated with the 21\,cm-line emitted and absorbed 
by spin-flips within the HI system. For this line to 
propagate, a fine-tuning of the brightness temperature $T_B$ 
of the cloud would be needed, see Fig.\,\ref{Fig-lowmom}, 
because photons that are absent at wavelengths $\sim$1\,cm 
would be required at the wavelength 21\,cm to maintain the 
thermal equilibrium in the spin system. The question whether or not thermal 
equilibrium is realized in the latter is, however, 
not directly accessible to observation: While $T_B$ is directly 
observable by a determination of the distance of non-illuminated cloud regions and 
the emitted intensity of the 21\,cm-line the determination of 
$T_S$ hinges on assumptions on the optical depth and various other 
brightness temperatures \cite{Dickey}. 

The astrophysical origin of the structure GSH139-03-69 appears 
to be a mystery. The point we are able to make here is that 
once such a cloud of HI particles has formed it likely 
remains in this state for a long period of time.  


\section{Two-loop corrections to the pressure revisited\label{Pres2}}

\subsection{General considerations}

In \cite{HerbstHofmannRohrer2004} the two loop corrections to the 
pressure of an SU(2) Yang-Mills theory in its deconfining phase 
were calculated omitting the term $\propto u_\mu u_\nu$ 
in Eq.\,(\ref{TLMprop}). Here we take this term into 
account in our calculation. 
\begin{figure}[htb]
\begin{center}
\epsfig{file=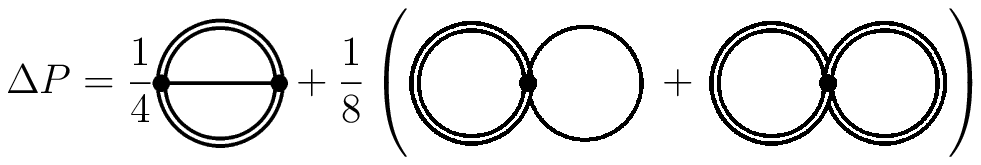,height=2.2cm,width=14cm}
\end{center}
\caption{Two-loop corrections to the pressure.\label{Fig-6}}
\label{2pressure}
\end{figure}
The two-loop corrections to the pressure are calculated as indicated in Fig.\,\ref{2pressure}. 
One has
\begin{equation}
\Delta P = \Delta P_{\tiny\mbox{nonlocal}} + \Delta P_{\tiny\mbox{local}}\,.
\end{equation}
The analytical expressions take the form
\begin{equation}
\label{Plocal}
\Delta P_{\tiny\mbox{local}} = \frac{1}{8i}\int \frac{d^4k}{(2\pi)^4} \frac{d^4p}{(2\pi)^4} 
\Gamma^{\mu\nu\rho\sigma}_{[4]abcd} D_{\mu\nu,ab}(k) D_{\mu\nu,cd}(p)
\end{equation}
and
\begin{equation}
\label{Pnonlocal}
\begin{split}
\Delta P_{\tiny\mbox{nonlocal}} &= \frac{1}{8i}\int \frac{d^4k}{(2\pi)^4} \frac{d^4p}{(2\pi)^4} 
\Gamma^{\lambda\mu\nu}_{[3]abc}(p,k,-p-k) \Gamma^{\rho\sigma\tau}_{[3]rst}(-p,-k,p+k) \\
& \quad D_{\lambda\rho,ar}(p) D_{\mu\sigma,bs}(k) D_{\nu\tau,ct}(-p-k).
\end{split}
\end{equation}
$D_{\mu\nu,ab}$ stands for the appropriate TLH and TLM propagator. 
The calculation proceeds along the lines of \cite{HerbstHofmannRohrer2004}: Because of the constraint 
(\ref{constraint1}) and the fact that $e\ge 8.89$ as a result of the one-loop 
evolution \cite{Hofmann2005} the vacuum parts in the TLH propagators do not contribute. 
After summing over color indices and contracting Lorentz indices the $\delta$-functions associated with 
the thermal parts of the TLH propagators render the integration 
over the zero components of the loop momenta trivial. The remaining integrations of spatial momenta 
are performed in spherical coordinates. At this point both constraints (\ref{constraint1}) and 
(\ref{constraint2}) are used to determine the boundary conditions for 
these integrations.

\subsection{Local diagrams}

Let us first introduce a useful convention: Due to the split of propagators 
into vacuum and thermal contributions 
in Eqs.\,(\ref{TLHprop}) and (\ref{TLMprop}) combinations of thermal and vacuum
contributions of TLH and TLM propagators arise in Eqs.(\ref{Plocal}) and (\ref{Pnonlocal}). 
We will consider these contributions separately and denote them by 
\begin{eqnarray}
\label{nomen}
\Delta P^{XYZ}_{\alpha_X\beta_Y\gamma_Z}\ \ \ \ \ 
\mbox{and}\ \ \ \ \Delta P^{XY}_{\alpha_X\beta_Y}
\end{eqnarray}
for the nonlocal diagram and the local diagrams in Fig.\,\ref{Fig-6}, 
respectively. In Eq.\,(\ref{nomen}) capital roman letters take the values $H$ or $M$, 
indicating the propagator type (TLH/TLM), and the associated small greek 
letters take the values $v$ (vacuum) or $t$ (thermal) or $c$ (Coulomb, the term $\propto u_\mu u_\nu$ 
in Eq.\,(\ref{TLMprop})). 

The correction $\Delta P^{HH}_{tt}$ was computed exactly in 
\cite{HerbstHofmannRohrer2004}. (The contributions $\Delta P^{HH}_{vt}$ and 
$\Delta P^{HH}_{vv}$ vanish due to (\ref{constraint1}).) 
The correction omitted in \cite{HerbstHofmannRohrer2004} is $\Delta P^{HM}_{tc}$: 
\begin{equation}
\label{PHMtc}
\begin{split}
\Delta P^{HM}_{tc} \,=\,&\frac{e^2}{8} \int \frac{d^4k}{(2\pi)^4}\frac{d^4p}{(2\pi)^4}\\ 
&\,\Big[\epsilon_{fac}\epsilon_{fdb}(g^{\mu\sigma}g^{\nu\rho}-g^{\mu\nu}g^{\rho\sigma})
+\epsilon_{fad}\epsilon_{fbc}(g^{\mu\nu}g^{\rho\sigma}-g^{\mu\rho}g^{\nu\sigma})
\Big] \delta_{ab} \delta_{cd}\times \\
& \left(g_{\mu\nu} - \frac{p_\mu p_\nu}{m^2}\right) 2\pi\,\delta(p^2-m^2)\,n_B(|p_0|/T)
\,\frac{i\,u_\rho u_\sigma}{\textbf{k}^2}\,.
\end{split}
\end{equation}
\begin{figure}[htb]
\begin{center}
\epsfig{file=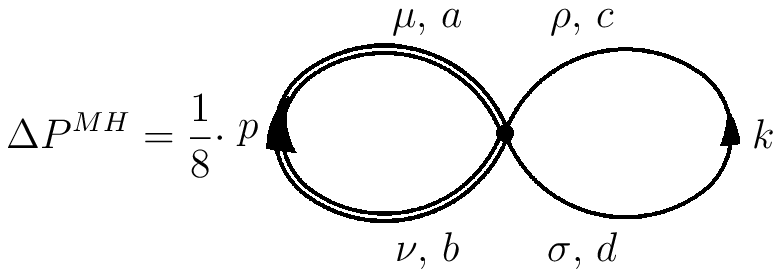,height=2.2cm,width=7cm}
\end{center}
\caption{The diagram for $\Delta P^{HM}$.\label{Fig-pmh}}
\end{figure}
Summing over color indices and contracting the Lorentz indices, see Fig.\,\ref{Fig-pmh}, yields
\begin{equation}
\label{PHMtcfinal}
\begin{split}
\Delta P^{HM}_{tc}\,=\,&\frac{e^2}{2} \int\frac{d^4k}{(2\pi)^4}\frac{d^4p}{(2\pi)^4}
\left(2+\frac{(p_0)^2}{m^2} \right)
2\pi\,\delta(p^2-m^2)\,n_B(|p_0|/T)\,\frac{i}{\textbf{k}^2}.
\end{split}
\end{equation}
Notice that $\Delta P^{HM}_{tc}$ is manifestly imaginary indicating that 
it must be canceled by the imaginary part of $\Delta P^{HM}_{tv}$ (The nonlocal 
diagram is manifestly real.). We are only able to estimate the modulus $|\Delta P^{HM}_{tc}|$. 
This is done in the following. 

Integrating over $p_0$ and introducing dimensionless variables as 
$x\equiv|\textbf{p}|/|\phi|$, $y\equiv|\textbf{k}|/|\phi|$, 
$\gamma\equiv k_0/|\phi|$, and $z\equiv\cos\theta\equiv\cos\angle(\textbf{p},\textbf{k})$, we have
\begin{equation}\label{PHMtcdimless}
\Delta P^{HM}_{tc}\,=\,\frac{i\,e^2\Lambda^4\lambda^{-2}}{2\,(2\pi)^5}
\sum_\pm\,\int dx\,dy\,dz\,d\gamma
\left(3+\frac{x^2}{4e^2}\right)
\frac{x^2\,n_B(2\pi\lambda^{-3/2}\sqrt{x^2+4e^2})}{\sqrt{x^2+4e^2}}\,
\end{equation}
where $\sum_\pm$ refers to the two possible signs of $p_0\rightarrow\pm\sqrt{\textbf{p}^2+m^2}$.
In dimensionless variables the constraints (\ref{constraint1}) 
and (\ref{constraint2}) read
\eab
\label{cond1}|k^2|\leq|\phi|^2 &\rightarrow& -1\leq\gamma^2-y^2\leq 1\\
\label{cond2}|(p+k)^2|\leq|\phi|^2 &\rightarrow& -1\leq 4e^2\pm2\sqrt{x^2+4e^2}\gamma-2x\,y\,z +\gamma ^2-y^2\leq 1\,.
\eae
On the one hand, to implement both conditions (\ref{cond1}) and (\ref{cond2}) exactly is technically very involved. 
On the other hand, neglecting condition (\ref{cond2}), as was done for $\Delta P^{HM}_{tv}$ 
in \cite{HerbstHofmannRohrer2004}, turns out to be insufficient for 
the correction $\Delta P^{HM}_{tc}$. Therefore, we fully 
consider (\ref{cond1}) and partly implement 
(\ref{cond2}) in our calculation. 

The integrand of Eq. (\ref{PHMtcdimless}) is positive definite. 
Condition (\ref{cond2}) represents a bound on a positive-curvature parabola 
in $\gamma$. Considering only the minimum $\gamma_\text{min}(x)=\mp\sqrt{x^2+4e^2}$ of this 
parabola, relaxes the restrictions on $x$, $y$, and $z$ meaning that the integration of the 
positive definite integrand is over a larger area than (\ref{cond2}) actually permits. Thus we obtain 
an upper bound for $\Big|\Delta P^{HM}_{tc}\Big|$. Replacing $\gamma$ with 
$\gamma_\text{min}$, condition (\ref{cond2}) reads
\begin{equation}
\label{newcond2}
-1\leq x^2+y^2+2x\,y\,z\equiv h(x,y,z)\leq1\,.
\end{equation}
Because this result is obtained for both signs of $p_0\rightarrow\pm\sqrt{\textbf{p}^2+m^2}$ we have $\sum_\pm=2$.
Let us now investigate the behavior of the function $h(x,y,z)$. Notice that $h(x,y,z)>-1$ 
because $h(x,y,-1)=(x-y)^2\ge 0$. The upper bound $h(x,y,z)\le 1$ puts restrictions on 
the upper limit $\min(1,z_+(x,y))$ for the $z$-integration where 
\begin{equation}
z_+(x,y)\equiv \frac{1-x^2-y^2}{2x\,y}\,.
\end{equation}
Hence $z$ runs within the range $-1\le z\le\min(1,z_+(x,y))$. The next task is 
to determine the range for $x$ and $y$ 
for which $-1\le z_+(x,y)\le 1$. Setting $z_+(x,y)>1$, we have
\eqb
\label{uppery}
0\le y<\tilde{y}(x)\equiv-x+1\,.
\eqe
Setting $z_+(x,y)>-1$, we have
\eqb
\label{luy}
0\le y_-(x)\equiv x-1<y<y_+(x)\equiv x+1\,.
\eqe
\begin{figure}
\begin{center}
\leavevmode
\leavevmode
\vspace{4.9cm}
\includegraphics{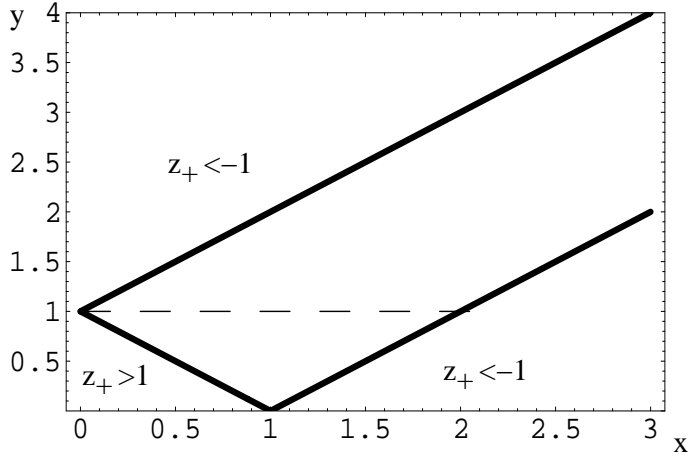}
\end{center}
\caption{\protect{\label{Fig-pmhL}} Admissible range for the $x$- and $y$-integration. The regions 
with $z_+<-1$ are forbidden, the dashed line represents the function $y=1$.}      
\end{figure}
The admissible range for $x$ and $y$ is depicted in Fig.\,\ref{Fig-pmhL}. 

\begin{figure}
\begin{center}
\leavevmode
\leavevmode
\vspace{4.9cm}
\includegraphics{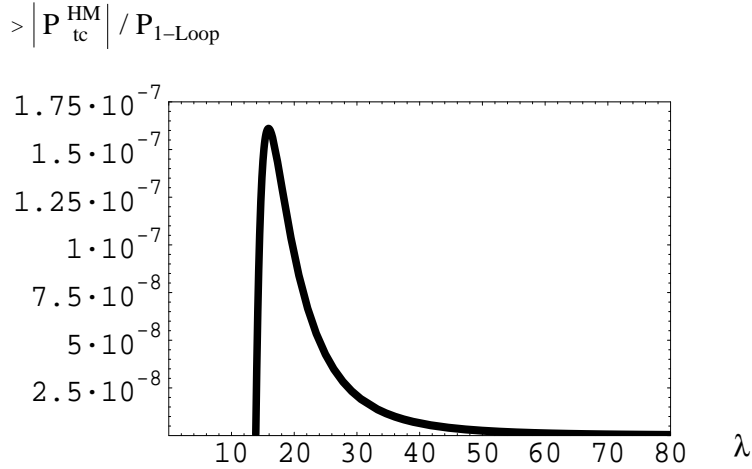}
\end{center}
\caption{\protect{\label{Fig-8}} Upper estimate for the 
modulus $\left|\Delta P^{HM}_{tc}\right|/P_{\tiny\mbox{1-loop}}$ 
as a function of $\lambda$ where $P_{\tiny\mbox{1-loop}}$ is the pressure due 
to one-loop fluctuations. (The ground-state part in $P$ is omitted in $P_{\tiny\mbox{1-loop}}$ \cite{Hofmann2005}.)}      
\end{figure}

To obtain limits on the $\gamma$-integration we solve condition (\ref{cond1}) for $\gamma$. 
For $y\ge 1$ we have
\eqb
\label{y>1}
\sqrt{y^2-1}\leq\gamma\leq\sqrt{y^2+1}\,\ \ \ \mbox{or} \ \ \ \ 
-\sqrt{y^2+1}\leq\gamma\leq-\sqrt{y^2-1}\,.
\eqe
For $0\le y<1$ we have
\eqb
\label{y<1}
-\sqrt{y^2+1}\leq\gamma\leq\sqrt{y^2+1}\,.
\eqe
Finally, we obtain:
\eab
\label{estmDPMH}
\left|\Delta P^{HM}_{tc}\right|&<&
\Big[\int_{0}^{1}dx\,\int_{0}^{\tilde{y}}dy\,\int_{-1}^{1}dz\int_{-\sqrt{y^2+1}}^{\sqrt{y^2+1}}d\gamma+
\int_{0}^{1}dx\,\int_{\tilde{y}}^{1}dy\,\int_{-1}^{z_+}dz\,\int_{-\sqrt{y^2+1}}^{\sqrt{y^2+1}}d\gamma+\nonumber\\ 
&&\int_{1}^{2}dx\,\int_{y_-}^{1}dy\,\int_{-1}^{z_+}dz\,
\int_{-\sqrt{y^2+1}}^{\sqrt{y^2+1}}d\gamma+2\int_{0}^{2}dx\,\int_{1}^{y_+}dy\,\int_{-1}^{z_+}dz\,
\int_{\sqrt{y^2-1}}^{\sqrt{y^2+1}}d\gamma+\nonumber\\ 
&&2\int_{2}^{\infty}dx\,\int_{y_-}^{y_+}dy\,\int_{-1}^{z_+}dz\,
\int_{\sqrt{y^2-1}}^{\sqrt{y^2+1}}d\gamma\Big]\nonumber\\ 
&&\frac{e^2\Lambda^4\lambda^{-2}}{(2\pi)^5}\left(3+\frac{x^2}{4e^2}\right)
\frac{x^2\,n_B(2\pi\lambda^{-3/2}\sqrt{x^2+4e^2})}{\sqrt{x^2+4e^2}}\,.
\eae
In Fig.\,\ref{Fig-8} a plot of the estimate for $\left|\Delta P^{HM}_{tc}\right|$ is shown as a 
function of $\lambda$. 

\begin{figure}
\begin{center}
\leavevmode
\leavevmode
\vspace{4.9cm}
\includegraphics{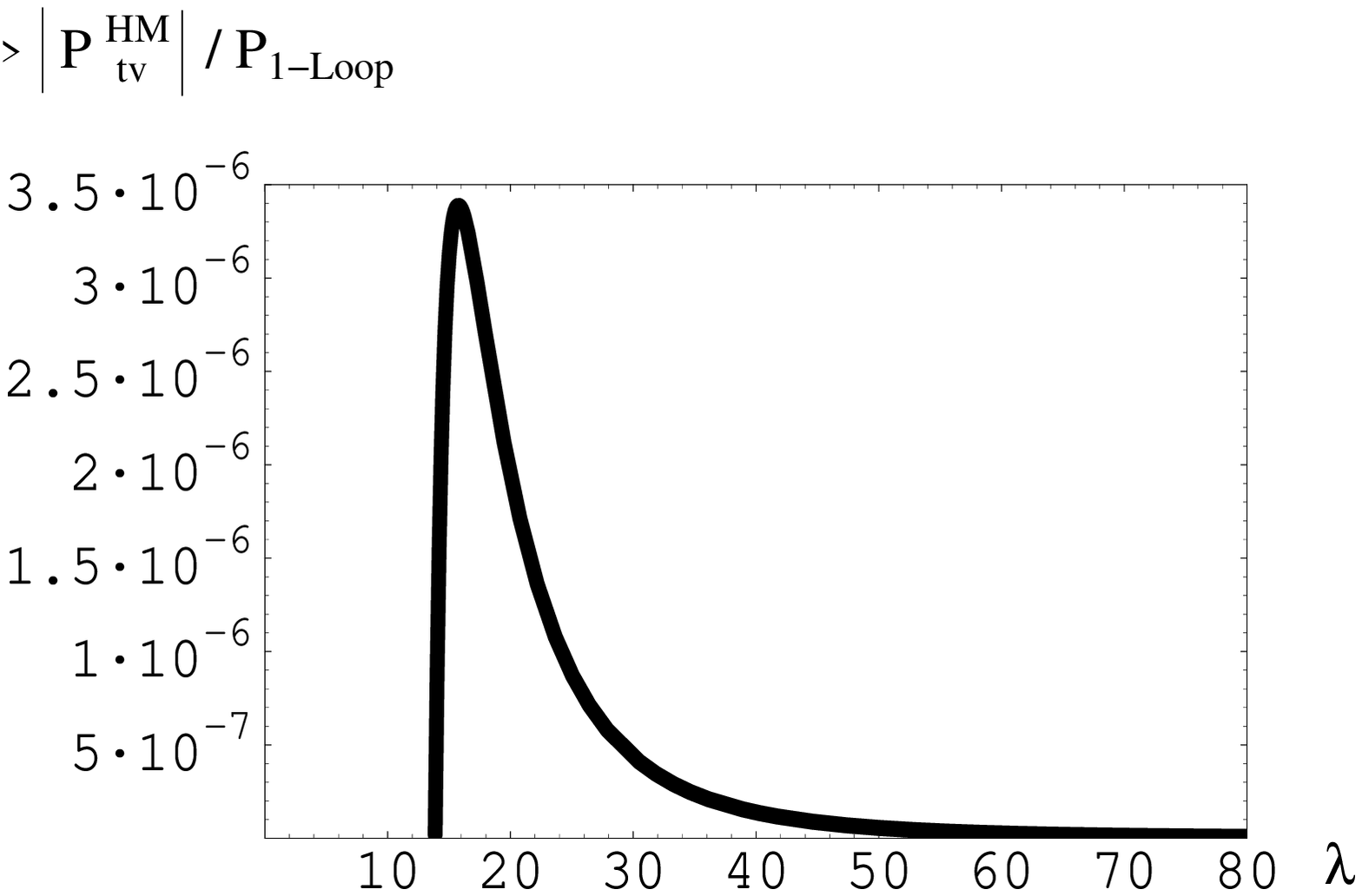}
\end{center}
\caption{\protect{\label{Fig-9}} Upper estimate for 
$\frac{\left|\mbox{Re}\,\Delta P^{HM}_{tv}\right|}{P_{\tiny\mbox{1-loop}}}$ as a function of $\lambda$.}      
\end{figure}

The next object to be considered is $\Delta P^{HM}_{tv}$. We know that $\Delta P^{HM}_{tc}$ is 
purely imaginary and thus is canceled by the imaginary part of $\Delta P^{HM}_{tv}$. 
Therefore the interesting quantity is $\mbox{Re}\,\Delta P^{HM}_{tv}$. We will be content with 
an upper estimate for $\left|\mbox{Re}\,\Delta P^{HM}_{tv}\right|$. This significantly 
improves the estimate for $|\Delta P^{HM}_{tv}|$ as it 
was obtained in \cite{HerbstHofmannRohrer2004} by a Euclidean rotation and by 
subsequently implementing the constraint (\ref{constraint1}) only. 
Here we would like to obtain a tighter estimate by also implementing 
(\ref{constraint2}) strictly along the lines developed for $\left|\Delta P^{HM}_{tc}\right|$. 
We have
\eab
\label{DeltaPHMtv}
\Delta P^{HM}_{tv}&=&-\frac{e^2}{2}\int \frac{d^4k}{(2\pi)^4}\,\frac{d^3p}{(2\pi)^3}\,dp_0\, 
n_B\left(\sqrt{\vec{p}^2+m^2}/T\right)\,\frac{i}{k^2+i\epsilon}\times\nonumber\\ 
&&\left(-4+\frac{\vec{p}^2}{m^2}-\frac{\left(\vec{p}\vec{k}\right)^2}{m^2\vec{k}^2}\right)
\frac{\delta\left(p_0-\sqrt{\vec{p}^2+m^2}\right)+\delta\left(p_0+\sqrt{\vec{p}^2+m^2}\right)}
{2\sqrt{\vec{p}^2+m^2}}\,.\nonumber\\
\eae
Going over to dimensionless variables, in our treatment of 
condition (\ref{cond2}) both choices for the sign of $p_0$ lead to one and the 
same constraint (\ref{newcond2}). Therefore the integral over the sum of $\delta$-functions in 
Eq.\,(\ref{DeltaPHMtv}) yields a factor of two. Notice that 
\eqb
\label{poleRe}
\lim_{\epsilon\to 0} \mbox{Re}\,\frac{i}{\gamma^2-y^2+i\epsilon}=\lim_{\epsilon\to 0} 
\frac{\epsilon}{(\gamma^2-y^2)^2+\epsilon^2}=\pi\,\delta(\gamma^2-y^2)
\eqe
and that the points $\gamma^2=y^2$ are not excluded by 
(\ref{cond1}). Performing the integrations over azimuthal angles and $\gamma$, we thus 
have
\eab
\label{|DeltaPHMtv|dimless}  
\left|\mbox{Re}\,\Delta P^{HM}_{tv}\right|&\le&\frac{e^2}{(2\pi)^4}\,\frac{\Lambda^4}{2\lambda^2}
\int dx\,dy\,dz\,\,x^2\,y\,\left|-4+\frac{x^2}{4e^2}-\frac{x^2\,z^2}{4e^2}\right|\,
\times\nonumber\\ 
&&\frac{n_B\left(2\pi\lambda^{-3/2}\sqrt{x^2+4e^2}\right)}{\sqrt{x^2+4e^2}}\,. 
\eae
Implementing constraint (\ref{cond2}) in the same way 
as in our estimate for $|\Delta P^{HM}_{tc}|$, yields the result depicted in Fig.\,\ref{Fig-9}.   
This estimate is about one order of magnitude better than the one 
for $\left|\Delta P^{HM}_{tv}\right|$ as obtained 
in \cite{HerbstHofmannRohrer2004}.

\subsection{Nonlocal diagram}

\begin{figure}[htb]
\begin{center}
\epsfig{file=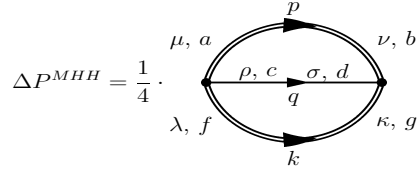,height=2.2cm,width=5.5cm}
\end{center}
\caption{The diagram for $\Delta P^{HHM}$.\label{Fig-phhm}}
\end{figure}

\begin{figure}
\begin{center}
\leavevmode
\leavevmode
\vspace{4.9cm}
\includegraphics{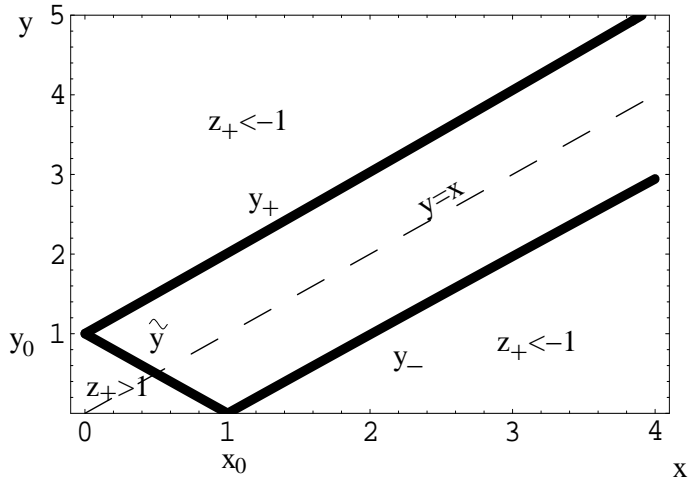}
\end{center}
\caption{\protect{\label{Fig-11}} The region of integration 
in the $x-y$ plane corresponding to Eq.\,(\ref{PHHM/ttcdimless}).}      
\end{figure}

\begin{figure}
\begin{center}
\leavevmode
\leavevmode
\vspace{5.9cm}
\includegraphics{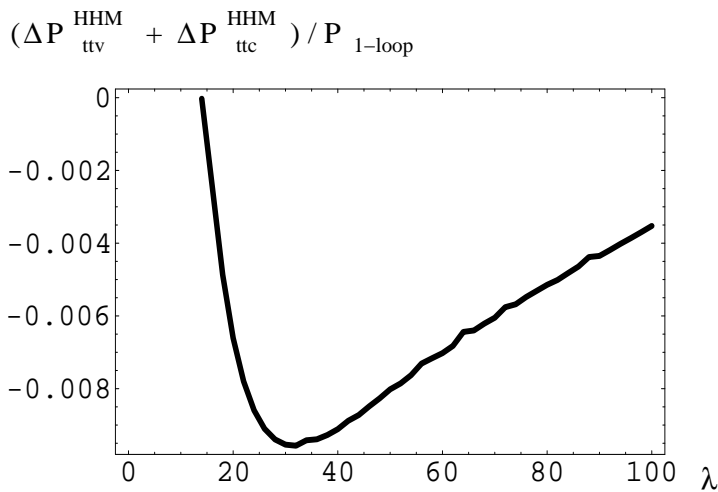}
\end{center}
\caption{\protect{\label{Fig-10} $\frac{\Delta P^{HHM}_{ttv}+\Delta P^{HHM}_{ttc}}{P_{\tiny\mbox{1-loop}}}$ 
as a function of $\lambda$.}}      
\end{figure}

Here we compute the correction $\Delta P^{HHM}_{ttv}+\Delta P^{HHM}_{ttc}$. In 
\cite{HerbstHofmannRohrer2004} the correction $\Delta P^{HHM}_{ttv}$ was 
computed exactly so that we can focus on $\Delta P^{HHM}_{ttc}$. The implementation of 
(\ref{constraint1}) is precisely as in \cite{HerbstHofmannRohrer2004}, only the 
integrand differs. Summing over color indices, see Fig.\,\ref{Fig-phhm}, we have  
\eab
\label{DeltaHHMttc}
\Delta P^{HHM}_{ttc}&=&-\frac{e^2}{4(2\pi)^6}\int\,d^4k\,d^4p\,d^4q\,
\delta(p+k+q)\,\frac{u_\rho u_\sigma}{\vec{q}^2}\times\nonumber\\ 
&&\Big[g^{\rho\mu}(p-q)^\lambda+g^{\lambda\rho}(q-k)^\mu+g^{\mu\lambda}(k-p)^\rho\Big]\times\nonumber\\
&&\left(g_{\mu\nu}-\frac{p_{\mu}p_{\nu}}{m^2}\right)\,\delta(p^2-m^2)
\,n_B\left(\frac{|p_0|}{T}\right)\times\nonumber\\ 
&&\Big[g^{\sigma\nu}(p-q)^\kappa+g^{\kappa\sigma}(q-k)^\nu+g^{\nu\kappa}(k-p)^\sigma\Big]\times\nonumber\\
&&\left(g_{\lambda\kappa}-\frac{k_{\lambda}k_{\kappa}}{m^2}\right)\,\delta(k^2-m^2)
\,n_B\left(\frac{|k_0|}{T}\right)\,.
\eae
Contracting the Lorentz indices 
is straight forward but 
cumbersome. The $q$-integration is trivial, and the 
$p_0$- and $k_0$-integration over the product 
of delta functions, $\delta(p^2-m^2)\,\delta(k^2-m^2)$, is 
performed as in \cite{HerbstHofmannRohrer2004}. Integrating over azimuthal angles, 
going over to dimensionless variables, and introducing the abbreviation $S(x,e)\equiv\sqrt{x^2+4e^2}$ 
yields
\eab
\label{PHHM/ttcdimless}
\Delta P^{HHM}_{ttc}&=&-\frac{e^2\Lambda^4\lambda^{-2}}{4(2\pi)^4}
 \int dx\,dy\,dz\, 			
 \frac{n_B\left(2\pi\lambda^{-3/2}S(x,e)\right)\,n_B\left(2\pi\lambda^{-3/2}S(y,e)\right)}
 {S(x,e)\,S(y,e)}\times\nonumber\\
 &&\frac{x^2y^2}{x^2+y^2+2x\,y\,z}\left[8\left(4e^2-
 \frac{(S(x,e)\,S(y,e)+x\,y\,z)^2}{4e^2}\right)\right.-\nonumber\\
 &&\frac{8e^2+x^2+y^2-2\,S(x,e)\,S(y,e)}{e^2}
 \left(S(x,e)\,S(y,e)+x\,y\,z\right)+\nonumber\\
 &&\frac{8e^2+x^2+y^2+2\,S(x,e)\,S(y,e)}{16e^4}\left(S(x,e)\,S(y,e)+x\,y\,z
 \right)^2-\nonumber\\
 &&\left. 2(8e^2+x^2+y^2-6S(x,e)\,S(y,e))\right].
\eae
As discussed in \cite{HerbstHofmannRohrer2004} the $z$-integration 
in Eq.\,(\ref{PHHM/ttcdimless}) is bounded as
\eqb
\label{zboundsP}
-1\le z\le \max\left(-1,\min(1,z_+(x,y))\right)
\eqe
where
\eqb
\label{z+Phhm}
z_+(x,y)\equiv\frac{1}{xy}\left(\frac{1}{2}+4e^2-S(x,e)\,S(y,e)\right)\,.
\eqe
The next task is 
to determine the range for $x$ and $y$ 
for which $-1\le z_+(x,y)\le 1$. Provided that $x<10,\,y<10$ one can solve the condition 
$z_+(x,y)>1$ for $y$. This yields  
\eqb
\label{yrange+1}
0\le y\le \tilde{y}(x)\equiv \frac{-(1+8e^2)x+\sqrt{1+16\,e^2}\,S(x,e)}{8e^2}\,.
\eqe
Notice that the intersection of $\tilde{y}(x)$ with the $y$- and $x$-axis is at 
$y_0=x_0=\sqrt{1+\frac{1}{16e^2}}\sim 1$. Thus our above assumption 
certainly is satisfied. Setting $z_+(x,y)>-1$, yields
\eab
\label{yrange-1}
y_-(x)&\equiv&\frac{(1+8e^2)x-\sqrt{1+16\,e^2}\,S(x,e)}{8e^2}\le y
\le y_+\equiv \frac{(1+8e^2)x+\sqrt{1+16\,e^2}\,S(x,e)}{8e^2}\,.\nonumber\\ 
\eae
Notice the factor $\frac{1}{x^2+y^2+2x\,y\,z}$ in the integrand of 
Eq.\,(\ref{PHHM/ttcdimless}). Upon $z$-integration this transforms into 
$\sim\left.\log(x^2+y^2+2x\,y\,z)\right|_{-1}^{\max(-1,\min(1,z_+(x,y)))}$. 
For $z=-1$ there is an integrable singularity at $x=y$ presenting a problem 
for the numerical $x$- and $y$-integration. To cope with it 
we cut out a small band of width $2\delta$ centered at $x=y$ and observe stabilization of 
the result for $\delta\to 0$. In Fig.\,\ref{Fig-11} 
the region of integration is depicted in the $x-y$ plane.   
The computation of $\frac{\Delta P^{HHM}_{ttv}+
\Delta P^{HHM}_{ttc}}{P_{\tiny\mbox{1-loop}}}(\lambda)$ is performed with a one-loop running coupling $e$. The result 
is shown in Fig.\,\ref{Fig-10}. 

\subsection{Summary and discussion}

When including the `propagation' of the 0-component of the TLM mode 
in the computation of the two-loop correction to the
pressure there are quantitative but not qualitative modifications. 
This is suggested by the fact that the static electric screening 
mass diverges \cite{Hofmann2005}. The nonlocal 
one in Fig.\,\ref{Fig-6} dominates the other contributions. Recall that the 
contribution $\frac{\Delta P^{HM}_{tc}}{P_{\tiny\mbox{1-loop}}}$ is purely imaginary with a 
very small modulus and thus is canceled by the imaginary 
part of $\frac{\Delta P^{HM}_{tv}}{P_{\tiny\mbox{1-loop}}}$. 
The modulus of the real part of the 
latter was estimated in an improved way without invoking a rotation to 
the Euclidean. It is interesting to discuss the (exact) result 
for $\frac{\Delta P^{HH}_{tt}}{P_{\tiny\mbox{1-loop}}}$ in a qualitative way.

\begin{figure}
\begin{center}
\leavevmode
\leavevmode
\vspace{5.5cm}
\includegraphics{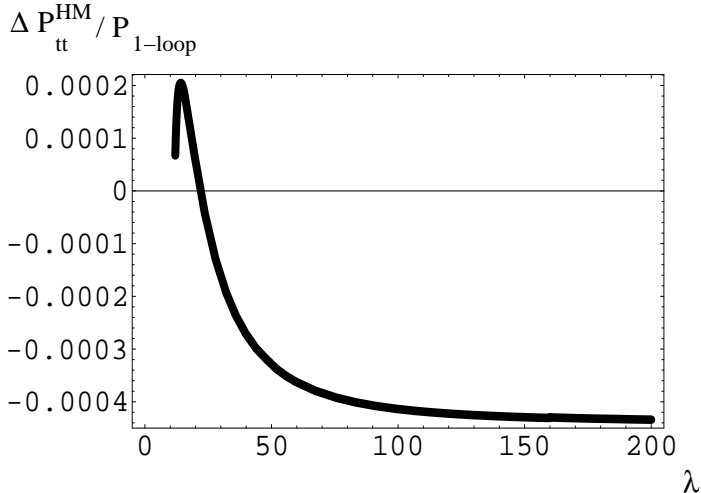}
\end{center}
\caption{\protect{\label{Fig-13} $\frac{\Delta P^{HM}_{tt}}{P_{\tiny\mbox{1-loop}}}$ 
as a function of $\lambda$.}}      
\end{figure}

As Fig.\,\ref{Fig-13} indicates, the contribution $\Delta P^{HM}_{tt}$ 
is negative and $\propto T^4$ since $P_{\tiny\mbox{1-loop}}\propto T^4$ for large $T$, 
see Fig.\,\ref{Fig-0B}. Let us propose an underlying scenario. First of all this effect is not 
due to isolated monopoles (M) and antimonopoles (A) as they are generated by the dissociation of 
large-holonomy calorons (repulsive M-A potential \cite{Diakonov2004}). 
(The probability of such processes is suppressed by 
$\exp\left[-\frac{m_M+m_A}{T}\right]\sim \exp[-8\pi^2]\sim 10^{-35}$ \cite{LeeLu1998,KraanVanBaalNPB1998,
vanBaalKraalPLB1998,Diakonov2004}.) The maximal distance between an M and its 
A, as generated by small-holonomy calorons (attractive M-A potential 
\cite{Diakonov2004}), is roughly given by the linear dimension 
$|\phi|^{-1}=\sqrt{\frac{2\pi T}{\Lambda^3}}$ 
of the coarse-graining volume. The typical on-shell TLM mode, however, 
carries a momentum $\sim T$. Thus for sufficiently large $T$ such a mode resolves the magnetic charges of the 
M or the A separately. If scattering of a TLM mode off of the M or the A within a small-holonomy 
caloron transfers a momentum larger than the typical 
binding energy $E_{\tiny\mbox{bind}}$ of the M-A system then 
isolated M and A are created. After screening, the masses of the latter are given 
by $\sim \frac{4\pi^2}{e}\,T$ \cite{Hofmann2005}: these particles decouple 
from the thermodynamics. 
\begin{figure}
\begin{center}
\leavevmode
\leavevmode
\vspace{4.3cm}
\includegraphics{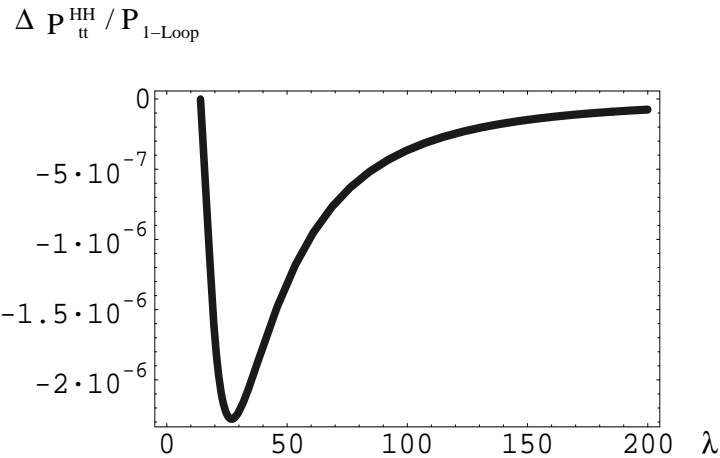}
\end{center}
\caption{\protect{\label{Fig-12} $\frac{\Delta P^{HH}_{tt}}{P_{\tiny\mbox{1-loop}}}$ 
as a function of $\lambda$.}}      
\end{figure}
The TLM mode responsible for this is less 
energetic after the scattering process, and thus its contribution to the thermodynamical pressure 
is diminished in comparison to the initial situation. From the observation 
that $\Delta P^{HM}_{tt}\propto T^4$ for $T\gg T_{c}$ we conclude 
that the typical $E_{\tiny\mbox{bind}}\propto T$. 

In general, the ground-state part of thermodynamical averages in the deconfining phase 
is perfectly saturated by the physics of small-holonomy calorons. There are, however, 
observables which are sensitive to the effects of large-holonomy calorons. 
In particular, such an observable is 
the spatial string tension $\sigma_s$, defined as
\eqb
\label{spss}
\sigma_s\equiv-\frac{\log\la \mbox{tr}\,{\cal P}\exp\left[ig\oint_C d\vec{l}
\cdot \vec{A}\right]\ra_T}{S(C)}\,,
\eqe
where ${\cal P}$ is the path-ordering symbol, $g$ denotes the fundamental coupling, 
and $S(C)$ denotes the minimal area enclosed by the contour $C$. The latter should 
be thought of as being the edge of a square 
of side-length $R$ in the limit $R\to\infty$. From lattice 
simulations one knows that $\sigma_s\propto T^2$ at high temperatures. Taking 
the logarithm of a thermal average, as it is done in Eq.\,(\ref{spss}), 
effectively singles out Boltzmann suppressed contributions. 
Due to the dissociation of large-holonomy calorons 
an integration over (nearly) static and screened M and A 
emerges in the partition function of microscopic SU(2) Yang-Mills 
thermodynamics \cite{Hofmann2005}. This integration is subject to a 
very small Boltzmann weight due to $m_{\tiny\mbox{M}}\sim m_{\tiny\mbox{A}}\sim \frac{4\pi^2}{e}\,T$: 
The associated part of the partition function $Z_{{\tiny\mbox{M+A}}}$ nearly 
can be factored out. This still holds true for the modified partition function 
when replacing $S_{YM}$ as $S_{YM}\to S_{YM}+
\log\left(\mbox{tr}\,{\cal P}\,\exp[ig\oint_C d\vec{l}\cdot \vec{A}]\right)$ 
which is relevant for the evaluation of the average under the 
logarithm in Eq.\,(\ref{spss}). It is the M-A factor in the modified partition 
function $Z^\prime$ that generates an area-law ($\propto R^2$), additive contribution 
to the numerator of the right-hand side of Eq.\,(\ref{spss}) 
\cite{GiovannangeliKorthals Altes2001}. The remaining factor in $Z^\prime$ describes the 
dynamics of fluctuating gauge fields and produces a perimeter-law ($\propto R$)
contribution. In the limit $R\to\infty$ only the M-A 
contribution survives in the expression for $\sigma_s$.
We maintain that all observables, 
which do not introduce external scales of resolution (the scale $|\phi|$ emerges in the absence of such scales!), 
such as the pressure, energy density, entropy density, polarization tensor with thermalized external momenta 
can be computed to any given accuracy in our approach (not only in principle but also practically).

\section{Summary and outlook} 

We have computed the one-loop polarization tensor for a massless 
mode with $p^2=0$ in the deconfining phase of thermalized 
SU(2) Yang-Mills theory. The result indicates that these modes 
do not propagate (strong screening) in a particular (temperature dependent) low-momentum range 
at temperatures, say, $2\,T_{c}\cdots 5\,T_{c}$. 

When applying these 
results to SU(2)$_{\tiny\mbox{CMB}}$ \cite{Hofmann2005,Hofmann20051,GiacosaHofmann2005}, 
identifying the massless mode with our photon, the prediction of a spectral 
gap in the low-frequency region of the black-body spectrum at $T\sim 2\,T_{\tiny\mbox{CMB}}\cdots 
5\,T_{\tiny\mbox{CMB}}$ emerges. Here $T_{\tiny\mbox{CMB}}=2.73\,$K. An experimental 
test of this prediction is proposed involving a cavity of linear 
dimension $\sim 50\,$cm. Such an experiment appears to be well feasible \cite{Morozova1993}. 

Another interesting aspect of our results on the polarization tensor is a possible explanation of 
the stability of cold, dilute, large, massive, and old innergalactic 
clouds of atomic hydrogen \cite{H1KB,Dickey,Dickey2001}. 
Namely, the mediation of the dipole interaction between hydrogen atoms, which 
is responsible for the formation of hydrogen
molecules, is switched off at distances on the scale of centimeters. Given the 
initial situation of an atomic gas of hydrogen with interparticle distance
$\sim 1\,$cm and brightness temperature $\sim 10\,$K, as reported in 
\cite{H1KB}, the formation of H$_2$ molecules is extremely 
suppressed as compared to the standard theory. 

In the remainder of the paper we have revisited the computation of the two-loop 
correction to the pressure first performed in \cite{HerbstHofmannRohrer2004}. 
On a qualitative level, our improved results agree with those in 
\cite{HerbstHofmannRohrer2004}. 

Our future activity concerning applications of SU(2) Yang-Mills thermodynamics will be 
focused on the physics of CMB fluctuations at low angular resolution.

\section*{Acknowledgements}
F. G. acknowledges financial support by the Virtual Institute VH-VI-041 
"Dense Hadronic Matter \& QCD Phase Transitions" of the Helmholtz Association.

\baselineskip25pt

\begin{thebibliography}{10}

\bibitem{Hofmann2005}
R. Hofmann, Int. J. Mod. Phys. A {\bf 20}, 4123 (2005).

\bibitem{HS1977}
B. J. Harrington and H. K. Shepard, Phys. Rev. D {\bf 17}, 105007 (1978).

\bibitem{'t HooftVeltmann}
G.~'t Hooft,
  Nucl.\ Phys.\ B {\bf 33} (1971) 173.
G.~'t Hooft and M.~J.~G.~Veltman,
  Nucl.\ Phys.\ B {\bf 44}, 189 (1972).
G.~'t Hooft,
  Int.\ J.\ Mod.\ Phys.\ A {\bf 20} (2005) 1336
  [arXiv:hep-th/0405032].

\bibitem{Nahm1984}
W. Nahm, Lect. Notes in Physics. 201, eds. G. Denaro, e.a. (1984) p. 189.

\bibitem{LeeLu1998}
K.-M. Lee and C.-H. Lu, Phys. Rev. D {\bf 58}, 025011 (1998).

\bibitem{KraanVanBaalNPB1998}
T. C. Kraan and P. van Baal, Nucl. Phys. B {\bf 533}, 627 (1998). 

\bibitem{vanBaalKraalPLB1998}
T. C. Kraan and P. van Baal, Phys. Lett. B {\bf 435}, 389 (1998).

\bibitem{Brower1998}
R. C. Brower, D. Chen, J. Negele, K. Orginos, and C-I Tan, Nucl. Phys. Proc. Suppl. {\bf 73}, 557 (1999).

\bibitem{Diakonov2004} 
D. Diakonov, N. Gromov, V. Petrov, and S. Slizovskiy, Phys. Rev. D {\bf 70}, 036003 (2004) 
[hep-th/0404042].

\bibitem{HerbstHofmann2004}
U. Herbst and R. Hofmann, hep-th/0411214.

\bibitem{HTL}
E. Braaten and R. D. Pisarski, Nucl. Phys. B {\sl 337}, 569 (1990).

\bibitem{HerbstHofmannRohrer2004}
U. Herbst, R. Hofmann, and J. Rohrer, Acta Phys. Pol. B {\bf 36}, 881 (2005).

\bibitem{Hofmann20051}
R. Hofmann, PoS JHW2005 {\bf 021} (2006) [hep-ph/0508176].

\bibitem{GiacosaHofmann2005}
F. Giacosa and R. Hofmann, hep-th/0512184. 

\bibitem{LandsmanWeert}
N. P. Landsman and C. G. van Weert, Phys. Rept. {\bf 145}, 141 (1987).

\bibitem{Morozova1993}
S. P. Morozova {\sl et al.}, Metrologia {\bf 30}, 369 (1993).

\bibitem{H1KB}
L. B. G. Knee and C. M. Brunt, Nature {\bf 412}, 308 (2001).

\bibitem{Dickey2001}
J. M. Dickey, Nature {\bf 412}, 282 (2001).

\bibitem{Dickey}
D. W. Kavars {\sl et al.}, Astrophys. J. {\bf 626}, 887 (2005).\\ 
D. W. Kavars and J. M. Dickey {\sl et al.}, Astrophys. J. {\bf 598}, 1048 (2003).\\ 
N. M. McClure-Griffiths {\sl et al.}, astro-ph/0503134. 

\bibitem{GiovannangeliKorthals Altes2001}
P. Giovannangeli and C. P. Korthals Altes, Talk given at 
19th International Symposium on Lattice Field Theory (Lattice 2001), Berlin, Germany, 19-24 Aug 2001, 
Nucl. Phys. Proc. Suppl. {\bf 106}, 616 (2002).


\end{thebibliography}
\end{document}